\documentstyle[12pt,aaspptwo,psfig]{article}
\def\etal{{\it et al. }}

\begin{document}

\title{On the Origin of Globular Clusters in Elliptical and cD Galaxies}

\author{Duncan A. Forbes}
\affil{Lick Observatory, University of California, Santa Cruz, CA 95064}
\and
\affil{School of Physics and Space Research, University of Birmingham,
Edgbaston, Birmingham B15 2TT, United Kingdom} 
\affil{Electronic mail: forbes@lick.ucsc.edu}

\author{Jean P. Brodie}
\affil{Lick Observatory, University of California, Santa Cruz, CA 95064}
\affil{Electronic mail: brodie@lick.ucsc.edu}

\author{Carl J. Grillmair}
\affil{Jet Propulsion Laboratory, 4800 Oak Grove Drive, Pasadena CA
91109} 
\affil{Electronic mail: carl@grandpa.jpl.nasa.gov}

\begin{abstract}

Perhaps the most noteworthy of recent 
findings in extragalactic globular cluster (GC) research are the 
multimodal GC metallicity distributions seen in massive early--type
galaxies. 
We explore the origin of these distinct GC populations, the implications
for galaxy formation and evolution, and identify several
new properties of GC systems. First, when we separate 
the metal--rich and metal--poor subpopulations, in galaxies with
bimodal GC metallicity distributions, we 
find that the mean metallicity of the metal--rich
GCs correlates well with parent galaxy luminosity but the mean
metallicity of the metal--poor
ones does not. {\it This indicates that the metal--rich GCs are closely
coupled to the galaxy and share a common chemical enrichment history
with the galaxy field stars}. The mean metallicity of the 
metal--poor population is largely independent of the galaxy
luminosity. 
Second, the slope of the GC system radial surface density 
varies considerably 
in early--type galaxies. However the galaxies with relatively populous GC
systems for their luminosity 
(called high specific frequency, S$_N$) {\it only} have shallow, extended
radial distributions. {\it A characteristic of high S$_N$ galaxies is
that their GCs are preferentially located in the outer galaxy regions
relative to the underlying starlight}. Third, we find that the ratio 
of metal--rich to metal--poor GCs correlates with S$_N$. In other
words, 
{\it high
S$_N$ galaxies have proportionately more metal--poor GCs, per unit galaxy
light, than low S$_N$ galaxies}. Fourth, 
{\it we find steeper
metallicity gradients in high S$_N$ galaxies.} This is due to the
greater number of metal--poor GCs at large galactocentric radii. 

We critically review current ideas for the origin of GCs in giant
elliptical (gE) and cD galaxies 
and conclude that the gaseous merger model of Ashman \& Zepf
(1992) is 
{\it unlikely} to account for the GC systems in these galaxies. 
Tidal stripping of the GCs from nearby galaxies appears to
contribute to the GC population in the outer parts of cD galaxies, but
we suggest that the 
vast majority of GCs in gE and cD galaxies have probably formed {\it
in situ}. 
We speculate that these indigenous GCs formed in 
two distinct phases of star formation from gas of differing
metallicity, giving rise to the bimodal GC metallicity distributions.  
The metal--poor GCs are 
formed at an early stage in the collapse of the
protogalactic cloud. The metal--rich GCs formed out of more enriched gas,
roughly contemporaneously with the galaxy stars. {\it In this sense 
the metal--rich GCs
in elliptical galaxies are the analog of the metal--poor halo GCs
in spirals}. The disk GCs in spirals may represent a third 
phase of this formation process.

\end{abstract}

\section{Introduction}

Globular clusters have long been recognized as `galactic fossils'
and like their counterparts in palaeontology they provide a unique tool 
for studying galaxy formation and evolution at early times. In
particular, 
they
may provide the best probe of chemical enrichment and star formation 
in the initial stages of galaxy formation. 
Observations of globular clusters (GCs) may enable us to distinguish between
variants of the collapse and merger models for elliptical 
galaxy formation, a distinction that has been difficult to 
make in the past (Kormendy 1990).

The first step in using GCs to probe galaxy formation is to understand the
formation of GCs themselves.
A variety of models 
(e.g. Fall \& Rees 1985; Murray \& Lin 1992; Kumai, Basu \& Fujimoto 
1993a; Harris \& Pudritz 1994; 
Vietri \& Pesce 1995) have sought to explain the formation mechanism for GCs. 
These models focus on gas cloud physics, including metallicity,
cooling and star formation rates, shocks etc. Although many of the 
early models have been ruled out, some of 
the more recent models may eventually be 
accepted as a prescription for GC formation. 
In this paper we take a different approach, and attempt to 
determine where and when GCs formed by examining the systemic
properties of GC systems. 
This may provide constraints for the further refinement of the 
GC and galaxy formation models. 

The number of GCs in an early--type galaxy scales roughly
with the parent galaxy's luminosity. A useful convention 
is to normalize the GC counts per absolute luminosity of
M$_V$ = --15 (Harris \& van den Bergh 1981). 
The resulting `specific frequency', S$_N = N \times
10^{0.4(M_V + 15)}$, is probably a measure of GC formation efficiency. 
When normalized in this way, S$_N$ $\sim$ 5 for
most early--type galaxies with magnitudes --19 $>$ M$_V$ $>$ --22 (e.g. 
Kissler--Patig
1996). However, for --22 $>$ M$_V$ $>$ --23.5 several 
galaxies have S$_N$ $\sim$ 15 i.e., a factor of three times as many
GCs per starlight as `normal' S$_N$ galaxies. 
Furthermore, these galaxies do not simply have higher global S$_N$
values but they show evidence for high S$_N$ values at
{\it all} galactocentric radii (McLaughlin, Harris \& Hanes 1994). 
Finding an explanation for the 
`high' S$_N$ GC systems has been described by McLaughlin \etal 
(1994) as ``the most outstanding problem in globular cluster
system research''. The origin of these GC systems has yet to be clearly
identified and indeed the question of whether they are anomalous or merely
represent the extreme of some distribution has been the subject of debate. 

Environmental effects on GC systems have been suspected for some time 
(e.g. cluster ellipticals have on average higher S$_N$ values than
field ellipticals; Harris 1991). However it was generally   
believed that there were no correlations between 
the properties of a GC system and the characteristics of the cluster
of galaxies in which the parent galaxy resided (McLaughlin \etal 1994). 
Initial evidence that such correlations did in fact exist was presented by
West \etal (1995). More recently,  
based on a larger and more homogeneous sample 
than was previously available, Blakeslee (1996) has shown that 
S$_N$ is correlated with cluster velocity
dispersion, the local density of bright galaxies and 
X--ray temperature. 
His sample, of 5 cDs and 14 giant ellipticals (gEs), cover the whole
range from  S$_N$ $\sim$4 to $\sim$10.
Therefore, 
although cD galaxies have undoubtly had more complex evolutionary histories 
than gEs, their GC systems may actually have had a similar origin. 
Blakeslee concluded that 
bright central galaxies with high S$_N$
values (these are usually but not always cDs) are not `anomalous' in the sense that
they have too many GCs for their luminosity. Rather, these galaxies are
`underluminous' for their clustercentric position
and the number of GCs more accurately reflects the local cluster
mass density. 
The fact that there is never more than one massive high S$_N$ galaxy per cluster
is consistent with these findings. 

In this paper we focus on the origin of GCs in massive early--type galaxies
(i.e. gE and cD galaxies). 
These galaxies contain the most populous GC systems; which are 
often the best--studied. 
Their GCs may have had a similar origin 
and the galaxies themselves share some common
properties e.g., the main bodies of cD galaxies have a 
similar fundamental--plane relation to that of elliptical galaxies
(Hoessel, Oegerle \& Schneider 1987). 
First, we present several recent
developments in the study of GC systems that provide 
important constraints on the origin of GCs. Second, we discuss
these results within the framework of various models proposed to explain 
the properties of GC systems of early--type galaxies. Third, we 
present our views and supporting data for what we believe are the most 
viable formation mechanisms (i.e. tidal stripping and {\it
in situ} GC formation). Fourth, we present a case study of the cD galaxy 
NGC 1399 and
the Fornax cluster of galaxies. Finally, we summarize our 
preferred scenario for the origin of GCs and 
make several predictions to test these ideas.

\section{Recent Developments in Extragalactic Globular Cluster
Studies}

In this section we highlight several aspects of GC systems that have
recently been discovered or confirmed. They provide important
constraints on any model for GC formation. We have taken distance
moduli and the number of GCs (N$_{GC}$) 
directly from the 
McMaster Catalog (Harris
1996). The one exception is NGC 1404, for which we take N$_{GC}$ from
the recent HST study of Forbes \etal (1997), i.e. N$_{GC}$ = 620 $\pm$
130. We believe this to be more accurate than the number quoted by
Harris (1996), i.e. 950 $\pm$ 140 (see also section 3.5). 
Extinction--corrected V magnitudes come from Faber \etal (1989) and
are generally within a few tenths of those given in the RC3. The one
exception here is NGC 3311, in which the values are V$_T^0$ = 10.14 (Faber
\etal 1989) and V$_T^0$ = 11.16 (RC3). In a photometric study of the Hydra
cluster, Hamabe (1993) finds V$_T^0$ = 10.75 which is roughly the
average of the above two values. Here we adopt the Hamabe value which
gives M$_V$ = --22.7. 
Most of the galaxies discussed in this section are luminous 
gE or cD galaxies. In this sense they are
generally boxy, slow rotators with shallow central cusps (e.g. Faber
\etal 1996; Carollo \etal 1997). 
We 
assume H$_{\circ}$ = 75 km s$^{-1}$
Mpc$^{-1}$.

\subsection{Multimodal Globular Cluster Metallicity Distributions}

Although the initial evidence for multimodal GC color (metallicity)
distributions in early--type galaxies was weak, there are 
now a handful of convincing cases (see Table 1 for a summary). 
The existence of multimodal GC metallicity distributions 
suggests that GCs have formed in distinct star formation episodes from
different metallicity gas; a
single star formation episode would give rise to a unimodal distribution. 
The observed metallicity distributions effectively rule out a simple
monolithic collapse and provide a strong constraint for any GC
formation model.  
The data for 11 gE/cD galaxies from the literature are 
summarized in Table 1. For NGC 1399 and NGC 4486
(M87) there have been claims for a trimodal metallicity distribution. 
In these cases we will refer to the metal--poor, intermediate
metallicity and metal--rich GC populations. 
The difference between the metallicity of the metal--rich and
metal--poor peaks has a mean value and statistical error of 
$\Delta$[Fe/H] = 1.0 $\pm$ 0.1 dex.

The metallicity distributions quoted in Table 1 are,  
strictly speaking, ones of color. The bimodal colors represent GC
populations that have different metallicities or ages, or some
combination of both. This is the well--known age--metallicity
degeneracy that affects the interpretation of optical colors. 
If we assume that the
two populations have the same metallicity and that 
the red population is very old (i.e. 15 Gyr), then 
for a typical color difference of $\Delta$(V--I) $\sim$ 0.2, 
the blue population is younger by $\sim$ 13 Gyr.  
On the other hand, for contemporaneous populations 
$\Delta$(V--I) $\sim$ 0.2 indicates that the red population is more
metal--rich than the blue one by $\sim$ 1
dex in metallicity.
The latter appears much more reasonable given, for example, our
knowledge of the Milky Way GC system. In the discussion below we 
will assume that GC
colors mostly reflect GC metallicity but in
reality there will probably be some component due to age
effects (see also Zepf 1996). For example, Chaboyer \etal 
(1992) have derived an
age--metallicity relation for Milky Way halo GCs in which
GCs with $\Delta$[Fe/H] = 1 dex are separated in age by $\sim$ 4 Gyrs. 
High signal-to-noise spectra of a sample of blue and red
GCs in an elliptical galaxy are needed to further investigate this issue.

If the characteristics of the GC system in 
NGC 4472 (Geisler, Lee \& Kim 1996) can be extended to 
other galaxies, then we can explain the 
long--standing observation that the mean metallicity of GCs is about 0.5 dex 
lower than field stars in gEs (e.g. Harris 1991). 
This then simply arises because the 
metal--rich GCs have a similar metallicity to the galaxy stars, and
the metal--poor GCs are about 1
dex more metal--poor. 
With a similar number of GCs in each subpopulation, the {\it mean} offset
between GCs and field stars 
is 0.5 dex, as is generally observed.
The lowest luminosity galaxy in Table 1 has M$_V$ =
--21. Less luminous ellipticals with M$_V$ $>$ --21 also show a metallicity 
offset between
GC and galaxy stars (Forbes \etal 1996).

In terms of the number of GCs observed with high quality CCD
photometry, the best studied GC system is that of NGC 4472 (M49) in
Virgo (Geisler \etal 1996). Although it is not a high S$_N$ galaxy
(S$_N$ = 5.5 $\pm$ 1.7), it is a giant elliptical (M$_V$ = --22.7) that is
somewhat more luminous than NGC 4486 (M87). Several interesting results
were found by Geisler \etal which may also hold true for other
ellipticals with bimodal metallicity distributions. The two
metallicity peaks in NGC 4472 are located at [Fe/H] = --1.25 and --0.05, so that
$\Delta$[Fe/H] = 1.2. They can be well represented by two Gaussians,
both with the same dispersion of 0.38 dex, with 
about 2/3 in the metal--poor peak and 1/3 in the metal--rich peak. 
(As given in Table 1, 
for a total population of 6300 this indicates there are about 
4200 metal--poor and 2100 metal--rich GCs in NGC 4472.) 
The overall radial 
metallicity gradient of the GC system is almost entirely due to the changing
relative mix of the two populations with radius. There is little or no
evidence for a metallicity gradient within either GC subpopulation
suggesting a largely dissipationless formation process (similar to the halo of
our Galaxy). The mean metallicity of the metal--rich population closely matches
the underlying field star metallicity over a wide range in galactocentric
radius. The metal--poor GCs are simply offset from the field stars by 1.2 dex. 
The metal--rich GCs are more centrally concentrated than the
metal--poor GCs, and have effective radii of 227$^{''}$ and
274$^{''}$ respectively. The galaxy itself has an effective radius of
about 100$^{''}$ (Bender, Burstein \& Faber 1992).

\subsection{The Globular Cluster Metallicity -- Galaxy Luminosity
Relation}

A relationship between GC mean metallicity and parent galaxy
luminosity (Z--L) was first suggested by van den Bergh (1975) and
later supported by the spectroscopic 
metallicity data of Brodie \& Huchra (1991).
The existence
of this relation, at least for low luminosity galaxies, was disputed
by Ashman \& Bird (1993). They claimed that the GC metallicity was
independent of galaxy luminosity. 
Larger galaxy samples (Durrell \etal 1996; Forbes
\etal 1996) have now confirmed that such a relationship 
of the form Z $\propto$ L$^{0.4}$, 
does indeed exist, over 10 magnitudes.   
The slope of the relation is the same as the
galaxy Z--L relation (Brodie \& Huchra 1991). Taken together these
results indicate
that GCs have had a similar chemical enrichment history to their
parent galaxy. The Z--L relation holds important clues about GC and
galaxy formation processes, but the large scatter, which
may be mostly observational error, makes it difficult to reach any
definitive conclusions. 

In the previous studies of the GC Z--L relation, the GC metallicity was
taken to be the mean of the entire GC system. We now know that in
several cases, the GC system is better represented by two GC
mean metallicities corresponding to the metal--rich and metal--poor
subpopulations. 
Although observational errors may dominate, there are other reasons for
expecting some scatter in the Z--L relation. 
In particular, some of the scatter will be due to changing
proportions of the metal--poor and metal--rich GC populations from one
galaxy to the next, which is further complicated by observations that
cover different galactocentric radii (metal--poor GCs are 
more radially extended than the metal--rich ones).
In Fig. 1 we plot the mean metallicity of the two subpopulations,
listed in Table 1, against the galaxy luminosity. We have assumed
errors of $\pm$ 0.1 for Washington color photometry, $\pm$ 0.15 for B--I and
$\pm$ 0.25 for V--I colors which reflect the scatter in the
color--metallicity relations.  
A weighted fit for the metal--rich GC  
peak gives [Fe/H] = --0.21 ($\pm$ 0.06) M$_V$ -- 4.6 ($\pm$1.4). 
The scatter of the metal--rich data points about the relation is less
than the estimated error bars, suggesting that our assumed metallicity
errors are overestimated. 
In the lower panel, for the metal--poor GCs, we have
simply offset the metal--rich relation downward by $\Delta$[Fe/H] = 1.0
dex (i.e. the average metallicity difference between the two peaks). 
There is considerably more scatter in the Z--L relation for the
metal--poor data points. This is unlikely to be caused by errors or
calibration problems and so the difference in scatter 
must be a real effect. Either there is
virtually no Z--L relation for metal--poor GCs or there is a `second
parameter effect' causing the significant scatter.  
Figure 1 clearly 
indicates that [Fe/H] for the metal--rich GCs is more closely coupled to
the luminosity of the galaxy than for the metal--poor GCs. 
This result will be
discussed further in section 3.4.  

\subsection{The Spatial Distribution of Globular Clusters}

It has been known for sometime that the radial distribution of GCs is 
often more extended than the starlight of the parent galaxy (see Harris
1991 for a review). The GC radial surface density profile is usually 
fit by:\\

\noindent
$\rho = \rho_o r^{\alpha}$ \hspace{5.5in} (1)\\

\noindent
where $\alpha$ is the radial slope. Kissler--Patig (1996) has
collected measurements of $\alpha$ from the literature. After excluding
the S0 galaxies, we show in Fig. 2 this slope versus the S$_N$ value
as given by Kissler--Patig.  
This figure shows that low S$_N$ galaxies have a wide range of GC
radial slopes, whereas high S$_N$ galaxies {\it only} have relatively
flat slopes. For S$_N$ $>$ 7 the mean slope is $\alpha$ = --1.43 $\pm$
0.12 (statistical error) and for S$_N$ $\le$ 7 it is $\alpha$ = --1.76
$\pm$ 0.09.  
There are no cases known of high S$_N$ galaxies with
steep GC radial slopes. This `zone of avoidance' is unlikely to be due
to selection effects as high S$_N$ galaxies have numerous GCs and so
are relatively easy to find and study. Thus another property of high
S$_N$ galaxies is that their GC systems have a very extended spatial 
distribution (see also Kaisler \etal 1996). 

A more useful quantity to examine would be the 
difference between the GC radial slope and the underlying starlight
slope. Unfortunately, the starlight slopes have only been measured over
similar galactocentric radii for a handful of galaxies. 
In the inner (i.e. r$^{1/4}$) regions of cD galaxies, the GC system is 
generally flatter than the galaxy starlight (Harris 1991; McLaughlin,
Harris \& Hanes 1993). 
In other words, a characteristic of high S$_N$ galaxies is that their 
GCs are preferentially located at large radii in the main body 
(r$^{1/4}$ part) of the galaxy.
In NGC 4486 (M87) the GC radial profile 
shows evidence for a change in slope at
the radius of the cD envelope (e.g. McLaughlin \etal 1993). This suggests
that the outer GCs and the envelope of cD galaxies have a similar origin.
As cD envelopes are generally thought to be a product of their cluster
environment (e.g. Kormendy \& Djorgovski 1989) this implies that GCs
in the envelope are also associated with some aspect of the cluster.

\section{Models for the Origin of Globular Clusters in gE and cD Galaxies}

Fall \& Rees (1988) characterized GC formation models into three
types -- primary, secondary and tertiary. In the primary formation
models, GCs may form well before the galaxy. 
In secondary formation models, GCs form at essentially 
the same time as the galaxy itself, and 
for tertiary formation models, GCs form after most of the parent
galaxy has formed. 
Given the numerous correlations between GCs and their parent galaxy (see
Harris 1991) 
and the absence of dark matter in GCs, 
formation well before the galaxy itself (e.g. Peebles \& Dicke
1968) and GC formation that is unrelated to galaxies (e.g. West 1993) 
seems to be ruled out. 
Theories for the 
formation of galaxies have traditionally been divided into the `collapse'
or the `merger' picture. The present majority view is that both collapse and
the merging of subunits (hierarchical clustering) play a role in the
formation and subsequent evolution of early--type galaxies. 
The presence of more than one population of GCs in massive early--type
galaxies effectively rules out a simple monolithic collapse. Either
the collapse was more complex (e.g. episodic) or some tertiary GC process 
has occurred (e.g. the accretion of existing GCs or the creation of
new GCs in a gaseous merger). 

Below we discuss specific models for the origin of GCs in
giant elliptical and cD galaxies:
1) a gaseous merger; 
2) cooling flows;
3) intracluster globular clusters;  
4) {\it in situ} GC formation 
and 5) tidal stripping of GCs from nearby galaxies. 
The merger model of Ashman \& Zepf (1992) and Zepf \& Ashman
(1993), makes detailed predictions for GC system formation on a
macroscopic level which are tested below. 
Most other models have not made such clear predictions and can therefore 
only be discussed on a more qualitative basis.

\subsection{The Gaseous Merger Model}

Van den Bergh's (1984) argued against the merger hypothesis 
on the grounds  
that ellipticals have many more GCs per unit light 
than spirals.
An additional problem is that a purely stellar merger of {\it two} 
ellipticals (in which no new GCs
are created) cannot explain the ratios of metal--poor to
metal--rich GCs. In particular,    
the GC metallicity -- galaxy luminosity
relation suggests that the metal--poor GCs would come from the lower luminosity
progenitor and would therefore also be fewer in number. Column 9 of
Table 1 shows that this is generally not the case, i.e. 
the number of metal--poor GCs 
exceeds the number of metal--rich GCs. 
The GC metallicity --
galaxy luminosity relation also suggests that metal--rich (luminous)
ellipticals are {\it not} made by simply merging several metal--poor
ellipticals. These issues and others have been discussed by van den
Bergh (1990). 

Schweizer (1987) has suggested that
new GCs are created from the gas of the merger progenitors. 
For a gaseous merger, the models of 
Kumai, Basu \& Fujimoto 
(1993a,b) propose that GCs form in the collision of
high velocity gas. Their models predict a mass--metallicity relation for
individual GCs which is not supported by the observational data (see
e.g. Forbes, Brodie \& Huchra 1996). 
Furthermore, this model can not apply to the GC systems of dwarf
galaxies which would probably be disrupted by high velocity
collisions. 

Another merger model that has
been the focus of much attention is that of Ashman \& Zepf (1992) and
Zepf \& Ashman (1993). Below we discuss this model in detail
in the context of GC formation in gE and cD galaxies. 
In their model the bimodal GC metallicity distributions are due to the
presence of metal--poor GCs from the progenitor spirals and
metal--rich ones created from the gas of the merged
galaxies. 
The metal--rich GCs are observed to be more
centrally concentrated than the metal--poor ones, as expected in their
model. 
One of the most noteworthy successes of this model is 
the discovery of protoglobular clusters in currently merging galaxies 
(e.g. Whitmore \etal 1993; Schweizer \etal 1996). 
The sizes, magnitudes, 
colors and mass distribution of the identified objects 
generally appear to be consistent with the properties of 
protoglobular clusters. Although some issues remain (such as
destruction of the very low mass objects) it seems plausible that these
objects will evolve into {\it bona fide} GCs. 
Thus the Ashman \& Zepf (1992) model appears consistent with the 
observations of GCs in currently merging galaxies.

Here we are interested in the origin of GCs in the most massive 
early--type galaxies. In the framework of the Ashman \& Zepf (1992)
hypothesis the gEs have been formed by 
massive gas--rich galaxies or several $\sim$ L$^{\ast}$ galaxies.
For cD galaxies they advocate multiple mergers which
in turn leads to several ``different epochs of 
[globular] cluster formation and
thus a spread in cluster ages''. 
Can a gaseous merger and subsequent GC formation, as described by
Ashman \& Zepf (1992), account for the
properties of GC systems in gE and cD galaxies ?

Previous studies of high S$_N$ galaxies (McLaughlin \etal 
1994) and
brightest cluster galaxies (Blakeslee 1996) have mentioned some
problems in trying to account for the populous GC systems with the
Ashman \& Zepf (1992) merger model. 
The case of NGC 3311 provides a particularly strong constraint for any
GC formation model as its GCs are relatively metal--rich with peaks at
[Fe/H] = --0.5
and +0.15 (Secker \etal 1995). 
It is hard to see how the GCs in a spiral
galaxy (which typically have mean [Fe/H] = --1.5) can account for the 
`metal--poor'
peak of [Fe/H] $\sim$ --0.5. 
Here we detail some of the observational data
which are contrary to the expectations of the Ashman \& Zepf (1992) 
merger model, 
in decreasing order of significance. \\

\noindent
$\bullet$ {\bf The metal--poor GCs are more numerous than the
metal--rich population}. In order to explain high S$_N$ galaxies 
by the merger of several normal S$_N$ galaxies, the majority of GCs
must be created in the merger event. This implies that the 
metal--rich GCs should be more numerous (by a factor of $\sim 3
\times$) than the
metal--poor GCs. This is not the case. 
In Table 1 we give the ratio of the number of metal--rich to
metal--poor GCs (this is equivalent to $\Delta$S = N$_{new}$/N$_{old}$
as defined by Ashman \& Zepf 1992). This ratio is expected to vary
from $\sim$ 1 for normal S$_N$ ellipticals to 4 for high S$_N$
galaxies. Table 1 shows that this ratio (column 9) is generally less
than 1. Furthermore, in the merger model 
this ratio should increase with increasing S$_N$
as larger numbers of new (metal--rich) GCs are formed. Instead we find
that it {\it decreases}, as shown in Fig. 3. 
Thus a property of high S$_N$ galaxies is that they have relatively 
more metal--poor GCs, in direct contradiction to that expected in the
Ashman \& Zepf (1992) model. For the cD galaxies in Table 1 
the number of metal--poor GCs significantly 
exceed the metal--rich ones. \\

\noindent
$\bullet$ {\bf The radial metallicity slope increases with
S$_N$}. Zepf \& Ashman (1993) predict that the ``colour [metallicity]
gradients of ellipticals with high S$_N$ values will be shallower than
those with normal S$_N$ values''. This would be so because in a high S$_N$
galaxy, large numbers of metal--rich GCs should be formed near the center and the
metal--poor GCs would be `pufffed up' to large galactocentric radii
(see Zepf \& Ashman 1993).  
We have collected metallicity gradients and S$_N$ values from the literature
and listed them in Table 2.  
In Fig. 4 we plot the [Fe/H] metallicity gradients versus S$_N$. 
Although there is considerable scatter,
the trend is for high S$_N$ ellipticals to have {\it steeper}
metallicity slopes than normal S$_N$ ellipticals. This trend is in
direct contradiction to the Zepf \& Ashman (1993) prediction and appears
to be a general problem for the merger model. 
Geisler \etal (1996) found that the 
merger model prediction 
failed to match the observed GC metallicity in both
absolute value and radial variation for NGC 4472.

We have made a crude prediction for the variation of metallicity
slope with increasing S$_N$ based on the data for NGC 4472 (Geisler
\etal 1996). As appears to be the case for NGC 4472 (M49), we will
assume that the observed GC metallicity gradients in ellipticals are due
solely to the changing relative mix of metal--rich and metal--poor GCs. 
Furthermore, we will assume that the only difference between a galaxy
of S$_N$ = 5.5 (i.e. NGC 4472) and one with S$_N$ = 11 is an increase in the 
number of metal--poor GCs (i.e. the number of metal--rich GCs remains
constant). 
Thus we can predict the metallicity slope of a galaxy with S$_N$ = 11. 
If these additional GCs are
distributed with the same surface density distribution 
as is currently seen, then we expect a 
metallicity gradient of $\Delta$[Fe/H]/$\Delta$logR = --0.58
dex/$^{''}$. 
The dashed line shown in Fig. 4 simply connects the data point for
NGC 4472 (i.e S$_N$ = 5.5 and $\Delta$[Fe/H]/$\Delta$logR = --0.41) to that
expected for an S$_N$ = 11 galaxy with a metallicity gradient of --0.58. It
reproduces the general trend surprisingly well. Note that the Ashman
\& Zepf (1992) predicted relation would have a gradient of the 
opposite sign to this
line. \\

\noindent
$\bullet$ {\bf High S$_N$ galaxies do not have peaky central GC
densities}. The gas in a merger quickly dissipates to the galaxy
center, where large numbers of new GCs should form. The expected
higher GC surface densities in the centers of high S$_N$ galaxies are 
{\it not} seen. 
In fact the surface density flattens off in
log space giving a GC system `core' (e.g. Grillmair, Pritchet \& van
den Bergh 1986; Lauer
\& Kormendy 1986; Grillmair \etal 1994a). 
The size of this core is larger in more 
luminous galaxies and this does not appear to be due to
destruction effects over time (Forbes \etal 1996). \\

\noindent
$\bullet$ {\bf Multiple mergers will not produce distinct 
metallicity peaks}. Based solely in luminosity terms, 
roughly 10 L$^{\ast}$ galaxies would be required
to form a cD galaxy or the more massive ellipticals, from multiple mergers. 
In general these galaxies will have
slightly different GC mean metallicities and the gas (from which new GCs
form) should be enriched by varying amounts.  
Therefore, unless all of the merging galaxies are similar, we would expect a 
broad, top--hat type distribution covering a range of metallicity, but instead 
distinct metallicity peaks are seen. In the case of NGC 4472 
(although not a cD it has a similar luminosity to a cD galaxy)
the metal--rich and metal--poor peaks have a similar dispersion when fitted with a
Gaussian. \\

\noindent
$\bullet$ {\bf Galaxies with the same luminosity should have similar 
S$_N$ values}. 
If galaxies are made from multiple mergers, why do some
apparently create GCs much more efficiently than others ? For example, NGC
4486 (M87)
and NGC 4472 (M49) both lie in the Virgo cluster and have essentially the
same luminosity, and yet they have quite different S$_N$ values (13 
and 5.5 respectively). In the merger hypothesis, this would indicate
that the GCs in NGC 4486 formed $\sim 3 \times$ more efficiently than
those in NGC 4472.\\

\noindent
$\bullet$ {\bf The GC surface density profiles of normal S$_N$ galaxies are not 
always flatter than the underlying starlight}. Zepf \& Ashman (1993)
``expect the globular cluster system to be noticeably flatter than the galaxy'' 
for normal (i.e. not high S$_N$) galaxies. As shown in Fig. 2, 
galaxies with 
S$_N$ $\le$ 7 have a wide range in GC surface density slopes
suggesting 
a wide range in the difference between GC and starlight slopes. 
Kissler--Patig \etal (1996) have studied several 
galaxies in the Fornax cluster with S$_N$ $<$ 7. 
All have GC slopes that are 
consistent with the starlight slope -- {\it none} are noticeably flatter 
contrary to the prediction of Zepf \& Ashman (1993). \\

\noindent
$\bullet$ {\bf The main body and envelope of cD galaxies appear to be
distinct}. For high S$_N$ galaxies, Zepf \& Ashman (1993) predict 
in general that ``the properties of the galaxy and the globular cluster system
should be similar''. Both the main body and the cD envelope should be
formed by the same mergers. In practice, 
the envelopes of cD galaxies are generally found
to be structurally distinct from the rest of the galaxy
(e.g. Mackie 1992). Furthermore, in both NGC 1399 and 
NGC 4486 (M87) the velocity
dispersion of the GCs is well in excess of the 
stellar velocity dispersion at large radii 
(Grillmair \etal 1994b; Huchra \& Brodie 1987; Mould \etal 1990); for
NGC 1399 the GC velocity dispersion in the cD envelope 
is close to that for the galaxies in the cluster.\\

To summarize, the gaseous merger model as proposed by Ashman \& Zepf (1992) 
for GC formation in massive early--type galaxies 
faces some serious difficulties. 
We note that some of these difficulties have been identified by 
Ashman \& Zepf (1997) and possible solutions suggested. However it is
our view that the disagreements between the observational data and the 
expectations from the Ashman \& Zepf merger model are still substantial. 
In particular, the model cannot explain the large numbers
of excess GCs present in high S$_N$ cD galaxies and appears to be in
conflict with GC trends in normal S$_N$ giant ellipticals.  

An open question is whether the Ashman \& Zepf (1992) model can
explain the origin of GCs in {\it low} luminosity ellipticals. 
It is currently not clear whether such galaxies are smaller versions
of gEs or if they form a distinct population (e.g. Carollo \etal
1997). 
Using the list of Kissler--Patig (1996), we find ellipticals with --19.5 $>$
M$_V$ $>$ --21.0 have S$_N$ = 4.6 $\pm$ 0.7. Whitmore \etal (1997)
show that in the best--studied merging galaxies the number of new GCs
created is equal to or less than the number of original GCs from the
progenitor spirals. This means that the S$_N$ values, after 15 Gyrs, 
are $\sim$ 2--3. Globular clusters need to form with much higher
efficiency in a gaseous merger if the S$_N$ values of merger remnants
are to match those of elliptical galaxies. 
So the argument that there are too many GCs in ellipticals to be
explained by mergers (van den Bergh 1984) may still be valid. 
If the GC systems of 
low luminosity ellipticals are not the result of a merger, then  
currently merging galaxies must form a 
galaxy with 
characteristics different from today's ellipticals (see also van 
den Bergh 1995).

\subsection{Cooling Flows}

It has been suggested that GCs could form in the material that cools
and condenses from cluster--wide hot gas (Fabian \etal 1984). 
Richer \etal (1993) proposed that the 
protoglobular clusters seen in NGC 1275  were the 
result of the $\dot{M} \sim 300 M_{\odot} yr^{-1}$
cooling flow. Subsequently, protoglobulars have been identified in
several galaxies without cooling flows (e.g. Whitmore \etal 1993, 1997; 
Schweizer \etal 1996). 
%Cooling flow GC formation has been extensively
%discussed by Grillmair \etal (1994b), 
%Harris, Pritchet \& McClure (1995) and Bridges \etal (1996a). 
%Quoting Bridges \etal (1996a), 
%the main argument against cooling flows as the dominant
%source of GC formation is that 
%``There is no correlation between cluster specific frequency S$_N$ or
%total number of clusters, and properties of the X--ray gas (X--ray
%luminosity, gas temperature, total gas mass, or cooling flow rate)''
%2) ``If cooling flows have been forming globular clusters continuously
%to the present day, we would expect to see some bright, blue clusters
%formed recently''. (Such clusters in NGC 1275 are most likely due to
%the merger, as no other cooling flow cD galaxy reveals bright blue clusters.)
%3) ``The metallicity of the X--ray gas is 3--5 times higher than the
%[Fe/H] $\sim$ --1 typical of globular clusters in gE/cD galaxies.'' 
%4) ``It is difficult to form objects with globular cluster masses from
%cooling flows.'' 
Bridges \etal (1996a) have discussed cooling flow GC formation and 
concluded that ``it seems most unlikely
that cooling flows are responsible for the high S$_N$
phenomenon''. Similar views were put forth by Grillmair \etal (1994b) and 
Harris, Pritchet \& McClure (1995). We refer the reader to these works
for further details. 

\subsection{Intracluster Globular Clusters}

West \etal (1995) proposed that there exists a population of GCs that
are associated with the potential well of the cluster of galaxies rather
than with any individual galaxy. A trend for the number of GCs to
correlate with X--ray temperature (once corrected for the 
distance of the galaxy from the dynamical center
of the cluster) was taken as evidence in favor of their
intracluster globular cluster (IGC) hypothesis. They suggested three
possible sources for these IGCs -- cooling flows, local density
enhancements and tidal stripping. Cooling flow GC formation has been
discussed above. Even if a 
mechanism could create GCs in 
local density enhancements outside of
galaxies, it would not account for the known correlations of GCs with
their parent galaxy (e.g. the GC metallicity--galaxy luminosity relation). 
Tidally stripped GCs may be more viable, especially in light of the
recent discoveries of intergalactic planetary nebulae (Theuns \&
Warren 1997). This is however 
unlikely to be the source of the entire $\sim$ 10,000
excess GCs required in high S$_N$ galaxies 
and metallicity is still an important constraint. 
We further discuss tidal stripping below.

\subsection{In Situ Formation}

Harris and co--workers (e.g. Harris 1991; McLaughlin \etal 1994; 
Harris, Pritchet \& McClure 1995; Durrell \etal 1996) advocate 
{\it in situ} formation for {\it most} of the GCs in gE 
and cD galaxies, with possibly a small
contribution from other sources.  
Durrell \etal (1996) found that GCs with M $\ge$ $10^5 M_{\odot}$
have a power--law mass distribution of the same 
form, which is remarkably independent of galaxy type, luminosity and 
environment.  
This suggests a similar GC formation mechanism for all galaxies,
including dwarfs, giants and cD galaxies.
The power--law is identical to that for 
giant molecular clouds (GMCs) in our Galaxy.
In a study of the currently merging galaxy NGC
3921, Schweizer \etal (1996) concluded that GMCs are
associated with the formation of protoglobular clusters. 
The largest GMCs may have similar masses and densities to the  
primordial fragments
proposed by Searle \& Zinn (1978) as the building blocks of galaxies
(see also Harris \& Pudritz 1994).
In this case a dwarf elliptical may represent a few Searle--Zinn
fragments whereas large galaxies are made up of numerous such fragments.

For cD galaxies, an argument in favor of {\it in situ} 
formation is that with  
a GC formation efficiency of $\le$1\% (Ashman \& Zepf
1992; Larson 1993) about 10$^{11} M_{\odot}$ of gas would be required 
to form the `excess' $\sim$10,000 GCs seen (assuming each GC has a 
mass of 10$^{5}
M_{\odot}$). 
Thus apparently most of the GCs in cD galaxies were formed at
an early epoch when the galaxy itself was in a mostly gaseous phase. 
In early--type galaxies, 
the observations that GCs are more metal--poor in the mean than the underlying
starlight at a given radius and that the GC systems tend to be more 
spatially extended than the starlight suggest that the GCs
formed before the bulk of the galaxy field stars (Harris 1991). The
various 
known  
correlations between the GC system and the parent galaxy suggest that 
the epoch of this `pre--galaxy' GC formation occurred just before the 
main collapse phase. 
But we now know, at least for the GC metallicity -- galaxy luminosity
relation (section 2.2), that the metal--poor GCs are poorly
correlated, if at all, with galaxy luminosity.
This relaxes the constraint on the epoch of formation for these GCs.  
A further modification to the standard collapse picture is required to explain
the presence of two distinct GCs populations (as indicated by the
now well--established bimodal color distributions) which rules out a
simple 
monolithic
collapse. Below we discuss our view for GC formation in a multiphase
collapse.  

At early times, low--metallicity 
pre--galaxy GCs and some field 
stars will form, enriching the protogalactic gas.  
If only a small fraction of the total gas is used in this initial
phase, then the remaining metal--enriched gas
should undergo further collapse. A second episode of star formation
is required to produce the metal--rich GCs. In this `galaxy' phase 
the vast majority of field stars also form. 
Such episodic formation in different metallicity gas 
could be expected to give rise to 
bimodal GC metallicity distributions. 
We would expect the pre--galaxy GCs to have a nearly
spherical distribution, show little or no rotation and have a large
velocity dispersion. There is some evidence to support this 
prediction. Grillmair \etal (1994b) found that a sample of
GCs in the outer regions of NGC 1399 
(which would be dominated by the metal--poor ones) had a
much higher velocity dispersion than the stars in the main body of NGC
1399 and with no measurable 
rotation. In NGC 4486, GCs also have higher velocity
dispersions than the galaxy stars in the outer regions (Huchra \& Brodie
1987; Mould \etal 1990). 
The metal--rich GCs formed in the second (galaxy) phase may show some rotation 
depending on the degree of dissipation during the collapse and the presence of
torques. 

We also expect the pre--galaxy GCs to have a high S$_N$ value associated
with them as very few field stars form in the initial phase. Thus high
S$_N$ galaxies may simply be those with a higher fraction of 
pre--galaxy (metal--poor) GCs. Some evidence to support this idea is shown in
Fig. 3 which shows that high S$_N$ galaxies have relatively more
metal--poor GCs than metal--rich ones. 
The physical mechanism for regulating the number of pre--galaxy GCs
may depend on local density. 
In general the star formation time scale 
varies as $\propto \rho^{-1}$, whereas the dynamical collapse time
varies as $\propto \rho^{-1/2}$. So if the local ambient density is
high, a larger fraction of the initial material turns into GCs 
leaving less gas for the second episode of GC 
formation. High S$_N$ galaxies could simply be those with more
locally dense and/or clumpy gas. Another possibility, is that the number of
pre--galaxy GCs depends on the total mass of the protogalactic
cloud. However we do not find a strong trend for the number of metal--poor
GCs (see Table 1) 
to scale with galaxy luminosity, suggesting that the mass of the
protocloud is not a major factor.

Perhaps the main 
problem with this collapse picture is the lack of a well--understood 
mechanism for initiating GC (and star) formation in 
the pre--galaxy phase, turning it off and
turning it on again at some later stage in the collapse.
Recently, Burkert \& Ruiz--Lapuente (1997) have proposed a model in
which, after the first star formation phase, subsequent star formation
may be stalled by several Gyrs. They propose that the hot gas
initially heated by SNe type II in the first phase, will only partially
cool before being re--heated by SNe type Ia. This re--heating delays
the second star formation phase. Their model was applied to small
galaxies; it would be very interesting if it could be extended to
include large galaxies.  
It is interesting that the metallicity difference
between the two GC populations is about 1 dex in [Fe/H] for most 
large galaxies. If confirmed by larger samples, this would 
suggest some underlying uniformity in the formation
process and the length of the dormant phase. If we crudely assume 
that all large galaxies have metal--poor GCs with a mean 
of [Fe/H] = --1.2 and
metal--rich ones with a a mean of [Fe/H] = --0.2, then the age--metallicity
relation for Milky Way halo GCs (Chaboyer \etal 1992) indicates 
that the metal--rich GCs formed $\sim$ 4 Gyrs after the metal--poor
ones. Some evidence for early star formation followed by an extended
dormant phase comes from a recent study of the Local 
Group Carina galaxy by Smecker--Hane \etal (1996). In this case the
first epoch of star formation occurred about 12 Gyrs ago and the
second phase $\sim$ 5 Gyrs later.

In the Milky Way, and by inference spiral galaxies in general, there are two 
distinct populations of GCs associated with the disk and halo components. 
With reference to the collapse picture above, the halo GCs 
have kinematics and metallicities similar to those of halo stars
(Carney 1993). 
This indicates that halo stars and GCs were formed in the halo 
at much the same time, and may therefore be the spiral galaxy analog to the 
metal--rich GC population in ellipticals. 
We speculate that the disk GCs in spirals represent a third stage of
the collapse process (which is absent in gE galaxies).

\subsection{Tidal Stripping}

The transfer of GCs from small galaxies to a high S$_N$ 
galaxy by tidal stripping has been suggested by several authors and
discussed by van den Bergh (1990). For
example, Forte \etal (1982) proposed that NGC 4486 (M87) had captured a
significant number of GCs from other Virgo cluster galaxies. The
exchange of GCs between galaxies has been modeled in a simple
way by Muzzio and
collaborators (e.g. Muzzio \etal 1984; Muzzio 1987). The main 
argument put forward against this hypothesis is that GCs and starlight
will be removed in the same proportions so that,  
although the total number of GCs is increased,  the S$_N$ value is
unaffected (van den Bergh 1984; Harris 1991). However there is an important
point to be made. The GCs in at least some galaxies tend to be more extended
radially than the underlying starlight so that the local
S$_N$ value increases with galactocentric radius (e.g. 
Forbes \etal 1996). In the case of NGC 4472, the global
S$_N$ is 5.5 whereas the local S$_N$ exceeds 30 at 90 kpc 
(McLaughlin \etal 1994). 
Thus GCs tidally stripped from the outer parts
of a galaxy may have a high S$_N$
value, similar to that of the cD envelope of M87 (S$_N$
$\sim$ 25). 
Tidal effects would, in their 
weakest form, remove only the outermost
GCs and if strong enough, the whole galaxy could be accreted by its
more massive neighbor. 
Fully--accreted galaxies will have their orbital energy converted into
internal heat, which acts to `puff--up' the outer envelope of the
primary galaxy (Hausman \& Ostriker 1978).

In Table 3 we list three nearby galaxies that may have undergone
a tidal interaction with their more massive neighbors. They are: NGC 1404 (E1)
located near NGC 1399 in 
Fornax; NGC 4486B (cE0) near NGC 4486 in Virgo and NGC 5846A (cE2) 
near NGC 5846
in a compact group. Faber (1973) suggested that the two compact
ellipticals were actually tightly bound cores of tidally stripped 
galaxies. 
This idea is supported by the simulations of Aguliar \& White (1986)
who show that tidal interaction results in a more
compact secondary galaxy.  
As the central velocity dispersion is largely unaffected by 
tidal stripping we can estimate the original luminosity of the
companion galaxy from the
L--$\sigma$ scaling relation of normal ellipticals (e.g. Faber \etal
1996). Furthermore using the GC metallicity--luminosity relation
from Forbes \etal (1996) we can estimate the mean metallicity of the
tidally stripped GCs. The predicted original luminosity of the galaxy 
and GC metallicity, along with
estimates of the uncertainty, are given in Table 3. The predicted 
luminosity, suggests that NGC 4486B has lost about $\sim$95\% of its stars
and NGC 5846A $\sim$70\%. On the other hand, NGC 1404 has an observed
luminosity that is consistent with its velocity dispersion, suggesting
that if tidal stripping has occurred it has removed very
little galaxy starlight (but it may still have stripped off outer GCs).

If some fraction of the 
GC system of the companion galaxies has been 
removed via tidal stripping, then perhaps 
these GCs can be identified by their metallicity 
in the more massive galaxy.
Figure 5 shows the metallicity distribution of GCs in NGC 1399, 4486 and
5846. We also show the predicted
metallicities of the acquired GCs (as
listed in Table 3) from the companion galaxies, i.e. NGC 1404, NGC
4486B and NGC 5846A respectively. 

For NGC 1404 the predicted GC mean metallicity is [Fe/H] =
--0.7 $\pm$ 0.3. 
Estimates of the number of GCs in NGC 1404 
have been made by 
Richtler \etal (1992) and Hanes \& Harris
(1986) from ground--based studies. 
For a distance modulus of 31.0, they found 
N$_{GC}$ = 880 $\pm$ 120 (S$_N$ = 3.2 $\pm$ 0.4) and 
N$_{GC}$ = 190 $\pm$ 80 (S$_N$ = 0.7 $\pm$ 0.3) respectively. 
Most recently Forbes \etal (1997), using HST data, calculated the number to be 
N$_{GC}$ = 620 $\pm$ 130 (S$_N$ = 2.3 $\pm$ 0.4). 
The Forbes \etal (1997) data had the advantages over ground--based
studies of 
very low contamination rates from foreground stars and background
galaxies, a 
background field and a pointing $\sim$ 10 arcmin from NGC 1399 on the
opposite side to NGC 1404 (allowing subtraction of GCs associated with
NGC 1399). We believe that the HST data offers a more reliable
estimate of the number of GCs and we will adopt this value (as given in
Table 1) which is close to the weighted average of the two
ground--based studies. 
%The weighted average of these is N$_{GC}$ = 700 $\pm$ 100 
%GCs with a global S$_N$ = 2.4 $\pm$ 0.4 
If we assume that NGC 1404 once had
a `normal' GC population with S$_N$ = 5 $\pm$ 1, then we can estimate
the number of missing GCs. Thus:\\

$N_{missing} = 5 \times 10^{-0.4(M_V + 15)} - N_{now}$ \hspace{8.5cm}
(2)\\

\noindent
For M$_V$ = --21.08, we estimate that NGC 1404 had an original GC
population of 1350 $\pm$ 270. 
So NGC 1404 may have lost $\sim$ 730 GCs with a mean metallicity of
[Fe/H] = --0.7 $\pm$ 0.3. On the otherhand, NGC 1399 reveals an  
intermediate metallicity peak at [Fe/H] $\sim$ --0.8 
(Ostrov \etal 1993), as shown in Fig. 5. 
The number of GCs with this intermediate metallicity is 
estimated to be 40\% of the total number of GCs (see Table 1),
i.e. about 2100 $\pm$ 700. 
It is plausible that a large fraction of
the intermediate metallicity population GCs in NGC 1399 have been
tidally stripped from NGC 1404. This possibility is explored further
in section 4 below. 

The GC metallicity distribution in NGC 4486, from Washington
photometry, is claimed to have three peaks
at [Fe/H] = --1.35,--0.8 and --0.1 (Lee \& Geisler 1993). However, we
note that central (Whitmore \etal 1995) 
and slightly off--center (Elson \& Santiago 1996) HST images reveal 
only the metal--poor and metal--rich peaks from V--I colors. 
Geisler \etal (1996) discussed this possible contradiction and
suggested that it could be due to the different radial samplings of
the three studies. Whatever the explanation, the presence of an
intermediate metallicity population is not crucial to our analysis,
only that there exist some 
fraction of GCs with [Fe/H] $\sim$ --0.8, which in turn could be those 
stripped from NGC 4486B. 
The predicted mean metallicity of the stripped GCs 
from NGC 4486B is [Fe/H] = --0.7 $\pm$ 0.3. 
The total number of GCs in the NGC 4486 system is 
13,000 $\pm$ 500 (Harris 1996). We estimate from the data of Lee \&
Geisler (1993) that the fraction of GCs  
at this intermediate metallicity is about 40\% or 
5200 $\pm$ 200. For a normal S$_N$ value of 5 $\pm$ 1 and
M$_V$ = --21.0 we would expect NGC 4486B to have an original 
population of 1250 $\pm$ 250 GCs. Thus NGC 4486B has not contributed
significantly to the overall GC system in NGC 4486 
but may represent some fraction of the intermediate metallicity GCs. 
Other galaxies may have also contributed GCs to NGC 4486.

The GC metallicity distribution in NGC 5846 has two broad 
peaks, the metal--poor peak ranging from 
--1.3 $<$ [Fe/H] $<$ --0.9 (Forbes, Brodie \& Huchra 1997). 
The metallicity of NGC 5846A GCs,
predicted from its original luminosity, 
is [Fe/H] = --0.8 $\pm$ 0.3. The total number of GCs in
NGC 5846 is estimated to be 3120 $\pm$ 1850 (Harris 1996). We note
that Forbes, Brodie \& Huchra (1997) estimate a total of 4670 $\pm$
1270 from HST imaging. Using the Harris value, there
are about 780 $\pm$ 463 metal--poor GCs. 
This compares with 800 $\pm$ 150 GCs
as the estimate of the original GC population in NGC 5846A. 
Thus tidally stripped GCs from NGC 5846A may have 
contributed to the metal--poor peak in NGC 5846.

Blakeslee (1996) found a correlation of S$_N$ with local galaxy
density for cluster galaxies, and noted that this 
can be understood in the context of tidal stripping. As 
the crossing time in high velocity 
dispersion clusters is shorter, individual galaxies can
pass close to the cluster center many times, increasing the 
occurrence of interactions accordingly.

\section{A Case Study: The Fornax Cluster}

The Fornax cluster of galaxies is a good region in which to search for
the effects of tidal stripping. The cluster itself is relatively
nearby at 16 Mpc and well--studied. Furthermore it is more
compact and perhaps dynamically older than the Virgo cluster. 
A sketch of the inner regions of the Fornax cluster is shown in
Fig. 6.  
The cD galaxy, NGC 1399, is centered in the cluster X--ray emission
which extends out to at least 38$^{'}$ or 175 kpc (Ikebe \etal 1996). 
The nearby elliptical NGC 1404 lies within the
X--ray envelope of NGC 1399 and appears to have an abnormally low
S$_N$ value of 2.3 $\pm$ 0.4. 
In section 4.5 we noted that NGC 1404 may have been tidally
stripped of GCs, which have since been acquired by NGC 1399 which has 
S$_N$ = 13 $\pm$ 4. 

%We now discuss further support for this idea with regard to NGC
%1399 and NGC 1404. 
Following the arguments of Faber (1973) we can make a crude estimate
of the tidal radius in NGC 1404, i.e. the radius beyond which material
has been tidally stripped by NGC 1399. The tidal radius $R$ in kpc is
given by:\\

$R = D (3.5 M_1 / M_2)^{-1/3}$ \hspace{4in} (3)\\

\noindent
where $D$ is the distance of closest approach, and the masses of NGC
1399 and NGC 1404 are denoted by $M_1$ and $M_2$ respectively. 
We will assume that $D$ is equal to the current projected separation
of 9.7$^{'}$ or 45 kpc (note that the orbital pericenter may be
somewhat smaller, and that the 
current physical separation is quite probably larger than 45 kpc). 
For M$_V$ = --21.45 and --21.08, L$_V$ = 3.2 and 2.3 $\times$
10$^{10}$ L$_{\odot}$ for NGC 1399 and NGC 1404 respectively. 
If both galaxies have the same M/L ratio then the tidal radius $R$ = 
15 kpc. There is evidence from X--ray observations and GC kinematics
that M/L $\sim$ 50--100 M$_{\odot}$/L$_{\odot}$ for NGC 1399
(Grillmair \etal 1994b). If NGC
1399 has a M/L some 10--20 times that of NGC 1404, then the tidal
radius becomes $R$ $\sim$ 10 kpc. Richtler \etal (1992) and Forbes
\etal (1997) estimate that the GC system of NGC 1404 terminates at a
galactocentric distance of about 18 kpc. 
This radius is similar to that estimated above and suggests 
that tidal stripping of the outer GCs is a plausible mechanism to
explain the absence of GCs beyond 18 kpc.

Globular cluster velocity dispersions also appear to match the
expectations from 
the tidal stripping hypothesis. 
Grillmair \etal (1994b) found that a sample of 50 GCs 
in NGC 1399 (with a mean galactocentric radius $\sim$ 40 kpc) 
had a velocity dispersion of nearly 400 km s$^{-1}$. This value is
much higher than the {\it stellar} velocity dispersion measured in
NGC 1399 at somewhat smaller radii, but is similar to the 
velocity dispersion measured for galaxies of the Fornax
cluster. The similarities in velocity dispersions of GCs 
and Fornax galaxies is naturally consistent with a tidal stripping
scenario within the dark matter halo of NGC 1399, 
since the material stripped from a passing or infalling
galaxy will retain most or all of the specific angular momentum (with
respect to NGC 1399) of its parent galaxy. Arnaboldi \etal (1994) find
a remarkably similar velocity dispersion among the planetary nebulae
at distances $\sim$ 25 kpc, increasing confidence in the GC 
result and lending support to the idea that tidal stripping 
contributes to both the GC system and stellar 
envelope of cD galaxies.

Have any other galaxies, 
besides NGC 1404, in the Fornax cluster been tidally stripped
of GCs which may also contribute to the GC system of NGC 1399 ? 
The GC systems of several Fornax cluster 
galaxies have been studied recently by
Kissler--Patig \etal (1996). Using (m--M) = 31.0, V band magnitudes from
Faber \etal (1989) and N$_{GC}$ from Kissler--Patig \etal we list
their S$_N$ values, luminosities, predicted mean GC metallicities and 
projected distances from NGC 1399 in Table 4. 
Table 4 seems to indicate that S$_N$ increases with distance from the
cluster center, although the formal errors are consistent with a
constant S$_N$ value. 
Nevertheless the trend with projected distance may indicate that
several Fornax galaxies have undergone some tidal stripping. 

Further evidence for tidal stripping comes from a comparison of the GC
surface density slope and that of the underlying starlight. Initially
a GC system may be more extended than the starlight with a flatter
slope, tidal stripping will tend to remove the outermost GCs so that
the GC system becomes truncated and 
the GC surface density slope may approach that of the stellar profile.
Fig. 2 shows that low to
normal S$_N$ galaxies have a wide range of GC radial slopes and
presumably a wide range in the difference between GC and starlight
radial profiles. 
Kissler--Patig \etal (1996) have measured
the radial slopes for NGC 1399 and four other Fornax galaxies. 
The latter galaxies (NGC 1427, 1374, 1379 and 1387) 
have GC slopes that are the same
as the underlying starlight within the errors, whereas 
NGC 1399, has a GC slope that is noticeably flatter than the
starlight.

In the context of the above discussion, there are many similarities
between the Fornax cluster and the Virgo cluster. In Virgo, the
central cD (NGC 4486) is also embedded within an extensive X--ray
envelope. The GCs in the outer regions have a higher mean velocity 
dispersion than the galaxy stars (Huchra \& Brodie 1987; Mould \etal
1990). However, there are also some 
noteable differences between NGC 4486 (M87) and NGC 1399. 
NGC 4486 has far fewer near neighboring galaxies and a 
much smaller cD envelope than NGC 1399. This combined with the higher
velocity dispersion and dynamical youth of the Virgo cluster will
lead to differences in the number of accreted galaxies (and their GCs) between
the two cD galaxies. 

\section{Summary, Speculations and Predictions}

After reviewing the current ideas on GCs in massive 
early--type galaxies, we
have come to the conclusion 
that most
GCs were formed {\it in situ} in two star formation episodes.

In cD galaxies there
may be an additional contribution from GCs that have been tidally
stripped from neighboring galaxies. 
This accreted population should have a high local S$_N$ value and 
may have played an important role in the development of the cD envelope. 
In the cases of NGC 1399 and M87
we have identified an intermediate metallicity population of GCs 
that may have been acquired via tidal stripping.  
To a much
lesser extent all galaxies are 
accreting small galaxies and their GCs. 
We discuss the Fornax cluster in some detail and suggest that several
galaxies near the central cD (NGC 1399) may have been tidally stripped of
GCs. Evidence for this includes currently low S$_N$ values, steep GC
radial distributions that match the starlight and the high velocity
dispersion in the outer NGC 1399 GCs.

The gaseous merger model for GC formation (Ashman \& Zepf 1992),
although generally consistent
with currently merging galaxies, does not adequately explain the 
observational data for GCs in the massive early--type galaxies. 
We find that the bimodal GC metallicity distributions do not 
match the merger model
expectations when examined in detail. 
In particular, {\it the GC systems of high S$_N$ galaxies do not have
more metal--rich GCs as would be expected in the merger model  
but instead have more metal--poor ones}. This means
that GC metallicity gradients are steeper in high S$_N$ galaxies,
contrary to the predictions of Ashman \& Zepf (1992).

We favor an early collapse picture for galaxy 
formation (see for example 
Eggen, Lynden--Bell \& Sandage 1962; Larson 1975; Gott 1977; 
Sandage 1990) that probably involves some chaotic merging of many
small subunits 
(Searle \& Zinn 1978; Katz 1992). We speculate that there are three 
main phases in the collapse process (which we call pre--galaxy, galaxy
and disk phases) each of which may have associated
GC formation. First, in the early stages of the collapse,
there occurs rapid star formation.
Only a small fraction of the available gas is converted into stars,
and most of these stars are located in GCs which formed from  
the chaotic merging of dense, gravitationally bound
clumps. These GCs have a high velocity dispersion, 
are distributed throughout the cloud volume and are metal--poor. The
ratio of stars in GCs to field stars is large (i.e. 
the specific frequency is high). 
There will be little or no radial abundance gradient in the GC system 
from this phase.  
Both the field stars and GCs provide 
enrichment for the remaining
gas. In the second phase, after 
the gas cloud has undergone further collapse, 
field star formation is preferred over star formation in
GCs (possibly due to more efficient cooling at higher metallicities). 
Both field stars and GCs will have essentially the same 
metallicity (i.e. relatively metal--rich). They will have 
some rotation depending on the degree of dissipation in
the collapse. 
Unconsumed gas will eventually settle into a disk--like structure
near the galaxy center. Thus the small disks (e.g. Scorza \&
Bender 1996)
and peaky stellar distributions (e.g. Forbes, Franx \& Illingworth
1995) seen in low luminosity ellipticals may indicate a prolonged
dissipative collapse in the less massive systems. 
The variation of galaxy properties from large to small ellipticals may
reflect a sequence of decreasing gas-to-star conversion efficiency 
(see also Bender, Burstein \& Faber 1992). We note that the main
difficulty with this scenario is the lack of detailed mechanism for
creating two distinct phases of GC formation from a single halo
collapse. 
 
Spiral galaxies may represent the extreme of this process 
with very inefficient conversion of gas into stars. Spirals will  
have virtually no pre--galaxy star formation, so that the
enrichment levels of the second phase of collapse are reduced leading 
to GCs that
are in general more metal--poor (i.e. [Fe/H] $\sim$ --1.5) 
than the second phase GCs in ellipticals. Although they have quite
different metallicities, halo GCs in spirals appear to be the
analog to the metal--rich GCs in ellipticals. Spirals then go on
to form a prominent disk (and associated GCs) 
in the third phase of the collapse. 
The type of the final galaxy produced by the above process will depend
largely on the initial conditions of the the protogalactic gas cloud. 
For example, protoellipticals may be more dense, 
more clumpy and more chaotic than protospirals. 
As the ambient density
and clumpiness of the protoelliptical increases, more metal--poor GCs
with a high S$_N$ value are produced. At the extreme end of this
process, and when the protoelliptical is 
located close to the gravitational center of a
galaxy cluster, a cD galaxy is formed.

We conclude that episodic {\it in situ} GC formation and a contribution from
accretion/tidal stripping in the outer parts of 
cD galaxies may offer the `best bet' for solving the 
``most outstanding problem in globular
cluster system research'' (McLaughlin \etal 1994). 
This origin for GCs suggests several testable predictions. They include:\\

\noindent
$\bullet$ We expect the metal--poor GCs in early--type galaxies to form 
non--rotating, hot systems (high velocity dispersion) if their origin
is either from the pre--galaxy phase or from tidal stripping. The outer 
(metal--poor) GCs in both NGC 1399 and NGC 4486 (M87) have high
velocity dispersions. Future large samples, will hopefully be able to
determine GC kinematics as a function of metallicity. The metal--poor
GCs will be near spherically distributed.\\

\noindent
$\bullet$
We expect the metal--rich GCs in early--type galaxies to be 
slowly rotating systems, with more rotation seen in low luminosity
ellipticals than the giants. 
Some tentative evidence for GC rotation comes
from Hui \etal (1995) who noted that the metal--rich GCs in NGC 5128 
appeared to
share the rotation properties of the system of planetary nebulae. The
metal--rich GCs will closely match the ellipticity of the underlying galaxy.\\

\noindent
$\bullet$ We have found that the metallicity of the metal--rich
GC population is more closely coupled to parent galaxy luminosity than
the metal--poor population. This suggests that the GC mean 
metallicity -- galaxy luminosity relation is driven by the metal--rich
population, and galaxies with a significant metal--poor
population (i.e. the high S$_N$ galaxies) will lie below the mean
relation, to lower metallicity, for a given luminosity.\\

\noindent
$\bullet$ As more bimodal GC systems are discovered in ellipticals, 
we predict that
they will follow the current trends of proportionately 
more metal--poor GCs
and steepening radial metallicity gradients with increasing
specific frequency.\\

\noindent
$\bullet$ We expect low luminosity, low S$_N$ elliptical 
galaxies to have relatively more metal--rich than metal--poor GCs. 
Given the 
smaller number of GCs present (and an even smaller population of 
metal--poor ones, if present at all) it may be very difficult to 
detect a bimodal metallicity distribution. We
predict that many ellipticals, without obvious bimodality, 
will have metallicity (color)
distributions that are intrinsically broad, i.e. broader than the
photometric or spectroscopic measurement errors. \\

\noindent
$\bullet$ Tidally stripped galaxies may have steeper GC radial surface density 
slopes than non--stripped galaxies.
%and are similar to that of the underlying galaxy. 
Some initial
support for this prediction comes from Fleming \etal (1995) who finds
that galaxies in clusters have steeper GC radial slopes, at a given
magnitude, than those in low density environments. The study of
Fornax cluster galaxies by Kissler--Patig \etal (1996) also supports
this prediction.\\

\noindent
$\bullet$ Tidally stripped galaxies will have lower than average S$_N$
values, i.e. $\le$ 5. In the case of NGC 4486B and NGC 5846A we
suggest that the vast majority of GCs and field stars have been
removed by tidal interaction with their more massive neighbor, and we
expect S$_N$ $\sim$ 1 for these two compact ellipticals. \\

\noindent
{\bf Acknowledgments}\\
We thank R. Elson, W. Harris, M. Kissler--Patig, B. Whitmore, A. Zabludoff and
S. Zepf for useful discussions and comments. 
We also thank the referee, S. van den Bergh, for his careful reading
and suggestions for improving the paper. This
research was funded by the HST grant GO-05920.01-94A and GO-05990.01-94A\\

\newpage
\noindent{\bf References}

\noindent
%Ajhar, E. A., Blakeslee, J. P., \& Tonry, J. L. 1994, AJ, 108, 2087
%(ABT94)\\
Arnaboldi, M. \etal 1994, ESO Messenger, 76, 40\\
Ashman, K. M., \& Bird, C. M. 1993, AJ, 106, 2281\\
Ashman, K. M., \& Zepf, S. E. 1992, ApJ, 384, 50\\
Ashman, K. M., \& Zepf, S. E. 1997, Globular Cluster Systems,
(Cambridge University Press, Cambridge), in press\\
Aguliar, L., \& White, S. D. M. 1986, ApJ, 307, 97\\
%Aguliar, L., Hut, P., \& Ostriker, J. P. 1988, ApJ, 335, 720\\
%Ashman, K. M., Conti, A., \& Zepf, S. E. 1995, AJ, 110, 1164\\ 
%Baum, W. A., \etal 1995, AJ, 110, 2537\\
%Binggeli, B, Tammann, G. A., \& Sandage, A. 1987, AJ, 94, 251\\
%Beers, T. C., \& Geller, M. J. 1983, ApJ, 274, 491\\
%Bender, R. 1996 in New Light on Galaxy Evolution, edited by R. Bender
%and R. Davies, (Dordrecht, Kluwer), p. 181\\
Bender, R., Burstein, D., \& Faber, S. M. 1992, ApJ, 399, 462\\
%Bender, R. 1988, A \& A, 202, L5\\
%Bender, R. 1990, Dynamics and Interactions of Galaxies, p. 232, ed. 
%R. Wielen, Springer-Verlag, Berlin\\
%Bender, R., Dobereiner, S., \& Mollenhof, C. 1988, A \& AS, 74, 385\\ 
Blakeslee, J. P. 1996, Ph.D. Thesis, MIT\\
%Blakeslee, J. P., \& Tonry, J. L. 1996, ApJ, 465, L19\\
Bridges, T. A., Carter, D., Harris, W. E., \& Pritchet, C. J. 1996a,
MNRAS, 281, 1290\\
Bridges, T. A., Carter, D., Zepf, S. E., Ashman, K. M., Hanes, D. A.,
\& Dow, P. 1996b, Spectrum, 9, 8\\
Brodie, J. P., \& Huchra, J. 1991, ApJ, 379, 157\\
%Burrows, C., \etal 1993, Hubble Space Telescope Wide Field and
%Planetary Camera 2 Instrument Handbook, STScI\\
Burkert, A., \& Ruiz--Lapuente, P. 1997, ApJ, in press\\
Carollo, C. M., Franx, M., Illingworth, G. D., \& Forbes, D. A. 1997,
ApJ, in press\\
Carney, B. 1993, The Globular Cluster -- Galaxy Connection, 
ed. G. Smith and J. Brodie, (ASP conference series, San Francisco), p. 234\\
Chaboyer, B., Sarajedini, A., \& Demarque, P. 1992, ApJ, 394, 515\\
%Couture, J., Harris, W. E., \& Allwright, J. W. B., 1990, ApJS, 73,
%671\\
%Couture, J., Harris, W. E., \& Allwright, J. W. B., 1991, ApJ, 372, 97\\
%Davies, R. L., \etal 1987, ApJS, 64, 581\\
%Carollo, C. M., Danziger, I. J., \& Buson, L. 1993, MNRAS, 265, 553\\
Durrell, P. R., Harris, W. E., Geisler, D., \& Pudritz, R. E. 1996,
AJ, 112, 972\\
Eggen, O. J., Lynden--Bell, D., \& Sandage, A. 1962, ApJ, 136, 748\\
Elson, R. A. W., \& Santiago, B. X. 1996, MNRAS, 280, 971\\
%Forbes, D. A. 1991, MNRAS, 249, 779\\
Faber, S. M. 1973, ApJ, 179, 423\\
Faber, S. M., \etal 1989, ApJS, 69, 763\\
Faber, S. M., \etal 1996, preprint\\
Fabian, A. C., Nulsen, P. E. J., Canizares, C. R. 1984, Nature, 310, 733\\
Fall, S. M., \& Rees, M. J. 1985, ApJ, 298, 18\\
Fall, S. M., \& Rees, M. J. 1988, in Globular Cluster Systems in
Galaxies, edited by J. Grindlay and A. Philip (Reidel, Dordrecht), p. 323\\
Fleming, D. E. B., Harris, W. E., Pritchet, C. J., \& Hanes, D. A. 1995,
AJ, 109, 1044\\
%Forbes, D. A. 1994, AJ, 107, 2017\\
%Forbes, D. A., \& Thomson, R. C. 1992, MNRAS, 254, 723\\
%Forbes, D. A., Reitzel, D. B., \& Williger, G. M. 1994, AJ, in press\\
%Forbes, D. A., Franx, M., \& Illingworth, G. D. 1995, AJ, 109, 1988\\
%Forbes, D. A. 1996a, AJ, in press\\
%Forbes, D. A. 1996b, AJ, in press\\
Forbes, D. A., Brodie, J. P., \& Huchra, J. 1996, AJ, 112, 2448\\
Forbes, D. A., Brodie, J. P., \& Huchra, J. 1997, AJ, in press\\
%Forbes, D. A., Elson, R. A. W., Phillips, A. C., 
%Illingworth, G. D. \& Koo, D. C. 1994, ApJ, 437, L17\\
Forbes, D. A., Franx, M., \& Illingworth, G. D. 1995, AJ, 109, 1988\\ 
Forbes, D. A., Franx, M., Illingworth, G. D., \& Carollo, C. M. 1996,
ApJ, 467, 126\\
Forbes, D. A. \etal 1997, in preparation\\
%Forbes, D. A., Sparks, W. B., \& Macchetto, F. D. 1990, Paired and
%Interacting Galaxies, p. 431, ed. J. W. Sulentic, W. C. Keel and C.
%M. Telesco, NASA conference publication 3098\\
%Forte, J. C., Strom, S. E., \& Strom, K. M. 1981, ApJ, 245, L9\\
Forte, J. C., Martinez, R. E., \& Muzzio, J. C. 1982, 87, 1465\\
%Franx, M., \& Illingworth, G. D. 1988, ApJ, 327, L55\\
%Fleming, D. E. B., Harris, W. E., Pritchet, C. J., \& Hanes,
%D. A. 1995, AJ, 109, 1044\\
%Freeman, K. C. 1990, in Dynamics and Interactions of Galaxies, ed. 
%R. Wielen (Springer--Verlag, Berlin) p. 36\\
Geisler, D., Lee, M. G., \& Kim, E. 1996, AJ, in press\\
Gott, J. R. 1977, ARAA, 29, 299\\
%Goudfrooij, P., Norgaard Nielson, H. U., Hansen, L., Jorgensen, H. E.,
%\& de Jong, T. 1990, A \& A, 228, L9\\
%Goudfrooij, P., Hansen, L., Jorgensen, H. E., \& Norgaard Nielson, H. U.  
%1994, A \& AS, 105, 341\\
%Gunn, J. E. 1979, Active Galactic Nuclei, p. 213, ed.\ C.\ Hazard and
%S.\ Mitton, Cambridge University Press, Cambridge\\
Grillmair, C., Pritchet, C., \& van den Bergh, S. 1986, AJ, 91, 1328\\
Grillmair, C. \etal 1994a, AJ, 108, 102\\ 
Grillmair, C., \etal 1994b, ApJ, 422, L7\\ 
%Grillmair, C., \etal 1996, AJ 111, 2293\\ 
%Grillmair, C. J. 1995, personal communication\\
%Guzm\'an, R., Lucey, J. R., \& Bower, R. G. 1993, MNRAS, 265, 731\\
Hamabe, M. 1993, ApJS, 85, 249\\
%Hanes, D. A. 1977, Mem. RAS, 84, 45\\
Hanes, D. A., \& Harris, W. E. 1986, ApJ, 309, 599\\
Harris, G. L. H., Geisler, D., Harris, H. C., \& Hesser, J. E. 1992, AJ,
104, 613\\
%Harris, H. C., Baum, W. A., Hunter, D. A., \& Kreidel, T. J. 1991, AJ,
%101, 677\\ 
%Harris, W. E. 1986, AJ, 91, 822\\
%Harris, W. E. 1990, PASP, 102, 966\\
Harris, W. E. 1991, ARAA, 29, 543\\
%Harris, W. E. 1993, The Globular Cluster -- Galaxy Connection, p. 472,
%ed. G. Smith and J. Brodie, ASP conference series, San Francisco\\
Harris, W. E. 1996, www.physics.mcmaster.ca/Globular.html\\
%Harris, W. E. \etal 1986, AJ, 91, 822\\ 
%Harris, W. E., \& Hanes, D. A. 1987, AJ, 93, 1368\\
Harris, W. E., \& Pudritz, R. E. 1994, 429, 177\\ 
Harris, W. E., \& van den Bergh, S. 1981, AJ, 86, 1627\\
%Harris, W. E., Allwright, J. W. B., Pritchet, C. J., \& 
%van den Bergh, S. 1991, ApJS, 276, 491\\
%Huchra, J., Davis, M., Latham, D., \& Tonry, J. 1983, ApJS, 52, 89\\
Harris, W. E., Pritchet, C. J., \& McClure, R. D. 1995, ApJ, 441, 120\\
%Hau, G. K. T., \& Thomson, R. C. 1994, MNRAS, 270, L23\\
Hoessel, J. G., Oegerle, W. R., \& Schneider, D. P. 1987, AJ, 94, 1111\\
%Holtzman, J., \etal 1992, AJ, 103, 691\\
Hausman, M. A., \& Ostriker, J. P. 1978, ApJ, 224, 320\\
%Holtzman, J., \etal 1995a, PASP, in press\\
%Holtzman, J., \etal 1995b, PASP, submitted\\
Huchra, J., \& Brodie, J. P. 1987, AJ, 93, 779\\
Hui, X., Ford, H. C., Freeman, K. C., \& Dopita, M. A. 1995, ApJ, 449, 592\\
Ikebe, Y., \etal 1996, Nature, 379, 6564\\
%Illingworth, G. D., \& Franx, M. 1989, Dynamics of Dense Stellar
%Systems, p. 13, ed. D. Merritt, Cambridge University Press,
%Cambridge\\
%Jacoby, G. H., \etal 1992, PASP, 104, 599 (J92)\\
Kaisler, D., Harris, W. E., Crabtree, D. R., \& Richer, H. B. 1996,
AJ, 111, 2224\\
Katz, N. 1992, ApJ, 391, 502\\
Kissler--Patig, M. 1996, A \& A, in press\\
Kissler--Patig, M., Kohle, S., Hilker, M., Richtler, T., Infante, L.,
\& Quintana, H.  1996, A \& A, in press\\ 
Kormendy, J. 1990, in Dynamics and Interactions of Galaxies, ed. 
R. Wielen (Springer--Verlag, Berlin) p. 499\\
Kormendy, J., \& Djorgovski, S. 1989, ARAA, 27, 235\\
%Kormendy, J., \etal 1994, in ESO Workshop on Dwarf Galaxies, edited by
%G. Meylan (ESO, Garching) **\\
Kumai, Y., Basu, B., \& Fujimoto, M. 1993a, ApJ, 404, 29\\
Kumai, Y., Basu, B., \& Fujimoto, M. 1993a, ApJ, 416, 576\\
Larson, R. B. 1975, MNRAS, 173, 671\\
Larson, R. B. 1993 The Globular Cluster -- Galaxy Connection, 
ed. G. Smith and J. Brodie, (ASP conference series, San Francisco), p. 675\\
%Lauer, T. 1988, ApJ, 325, 49\\
Lauer, T., \& Kormendy, J. 1986, ApJ, 301, L1\\
Lee, M. G., \& Geisler, D. 1993, AJ, 106, 493\\
%Mould, J. R. \etal 1995, ApJ, 449, 413\\
Mackie, G. 1992, ApJ, 400, 65\\
%McLaughlin, D. E., \& Pudritz, R. 1996, ApJ, 457, 578\\
McLaughlin, D. E., Harris, W. E., \& Hanes, D. A. 1993, ApJ, 409,
L45\\
McLaughlin, D. E., Harris, W. E., \& Hanes, D. A. 1994, ApJ, 422,
486\\
Mould, J. R., Oke, J. B., De Zeeuw, P. T., \& Nemec, J. M. 1990 AJ,
99, 1823\\
Murray, S. D., \& Lin, D. N. C. 1992, ApJ, 400, 265\\
Muzzio, J. C. 1987, PASP, 99, 245\\
Muzzio, J. C., Martinez, R. E., \& Rabolli, M. 1984, ApJ, 285, 7\\
%Ostriker, J. P., \& Tremaine, S. D. 1975, ApJ, 202, L113\\
Ostrov, P., Geisler, D., \& Forte, J. C. 1993, AJ, 105, 1762\\
Peebles, P. J. E., \& Dicke, R. H. 1968, ApJ, 154, 891\\
%Perelmuter, J. L. 1995, ApJ, 454, 762\\
%Perelmuter, J. L., Brodie, J. P., \& J. P. 1995, AJ, 110, 620\\
%Quintana, H., \& Lawrie, D. W. 1982, AJ, 87, 1\\
Richer, H. B., Crabtree, D. R., Fabian, A. C., \& Lin, D. N. C. 1993,
AJ, 105, 877\\ 
Richtler, T., Grebel, E. K., Domgorgen, H., Hilker, M., \& Kissler,
M. 1992, A \& A, 264, 25\\
%Rosenblatt, E. I., Faber, S. M., \& Blumenthal, G. R. 1988, ApJ, 330, 191\\
Sandage, A. 1990, JRASC, 84, 70\\
%Sandage, A., \& Tammann, G. A. 1995, ApJ, 446, 1\\
Scorza, C., \& Bender, R. 1996 in New Light on Galaxy Evolution, 
edited by R. Bender and R. Davies, (Kluwer, Dordrecht), p. 55\\
Schweizer, F. 1987, Nearly Normal Galaxies, ed. S. Faber,
(Springer--Verlag, New York), p. 18\\ 
Schweizer, F., Miller, B. W., Whitmore, B. C., \& Fall, S. M. 1996,
AJ, in press\\ 
Searle, L., \& Zinn, R. 1978, ApJ, 225, 357\\
Secker, J., Giesler, D., McLaughlin, D. E., Harris, W. E. 1995, AJ,
109, 1019\\
%Secker, J. 1992, AJ, 104, 1472\\
%Secker, J., \& Harris, W. E. 1993, AJ, 105, 1358 (SH93)\\
Smecker--Hane, T. A., Stetson, P. B., Hesser, J. E., \& VandenBergh,
D. A. 1996, From Stars to Galaxies, ed C. Leitherer, U. Fritze von
Alvensleben and J. Huchra, (ASP Conference series, San Francisco), in press\\
%Stetson, P. B., 1987, PASP, 99, 191\\
%Surma, P. 1992, Structure, Dynamics and Chemical Evolution of
%Elliptical Galaxies, ed. I. J. Danziger, W. W. Zeilinger and K. Kjar,
%ESO: Garching, p. 669\\
%Tonry, J. L., Ajhar, E. A., \& Luppino, G. A. 1990, AJ, 100, 1416\\
%Tonry, J. L. 1991, ApJ, 373, L1\\
Theuns, T., \& Warren, S. J. 1997, MNRAS, 284, L11\\
van den Bergh, S. 1975, ARAA, 13, 217\\ 
van den Bergh, S. 1984, PASP, 96, 459\\
van den Bergh, S. 1990, Dynamics and Interactions of Galaxies, 
ed. R. Wielen, (Springer--Verlag, Berlin), p. 492\\
van den Bergh, S. 1995, AJ, 110, 2700\\
Vietri, M., \& Pesce, E. 1995, ApJ, 442, 618\\
%Weinberg, M. D. 1986, ApJ, 300, 93\\
West, M. J. 1993, MNRAS, 265, 755\\
West, M. J., Cote, P., Jones, C., Forman, W., \& Marzke, R. O. 1995,
ApJ, 453, L77\\
Whitmore, B. C., Schweizer, F., Leitherer, C. Borne, K., \& Robert,
C. 1993, AJ, 106, 1354\\
%Whitmore, B. C., \& Schweizer, F. 1995, AJ, 109, 960\\
Whitmore, B. C., Sparks, W. B., Lucas, R. A., Macchetto, F. D., \&
Biretta, J. A. 1995, ApJ, 454, L73\\
Whitmore, B. C., \etal 1997, in preparation\\
Zepf, S. E. 1996, Sant'Agata Conference on Interacting Galaxies, in press\\
Zepf, S. E., \& Ashman, K. M. 1993, MNRAS, 264, 611\\
Zepf, S. E., Ashman, K. M., \& Geisler, D. 1995, ApJ, 443, 570\\

%\noindent{\bf Figure Captions}

\begin{figure*}[p]
\centerline{\psfig{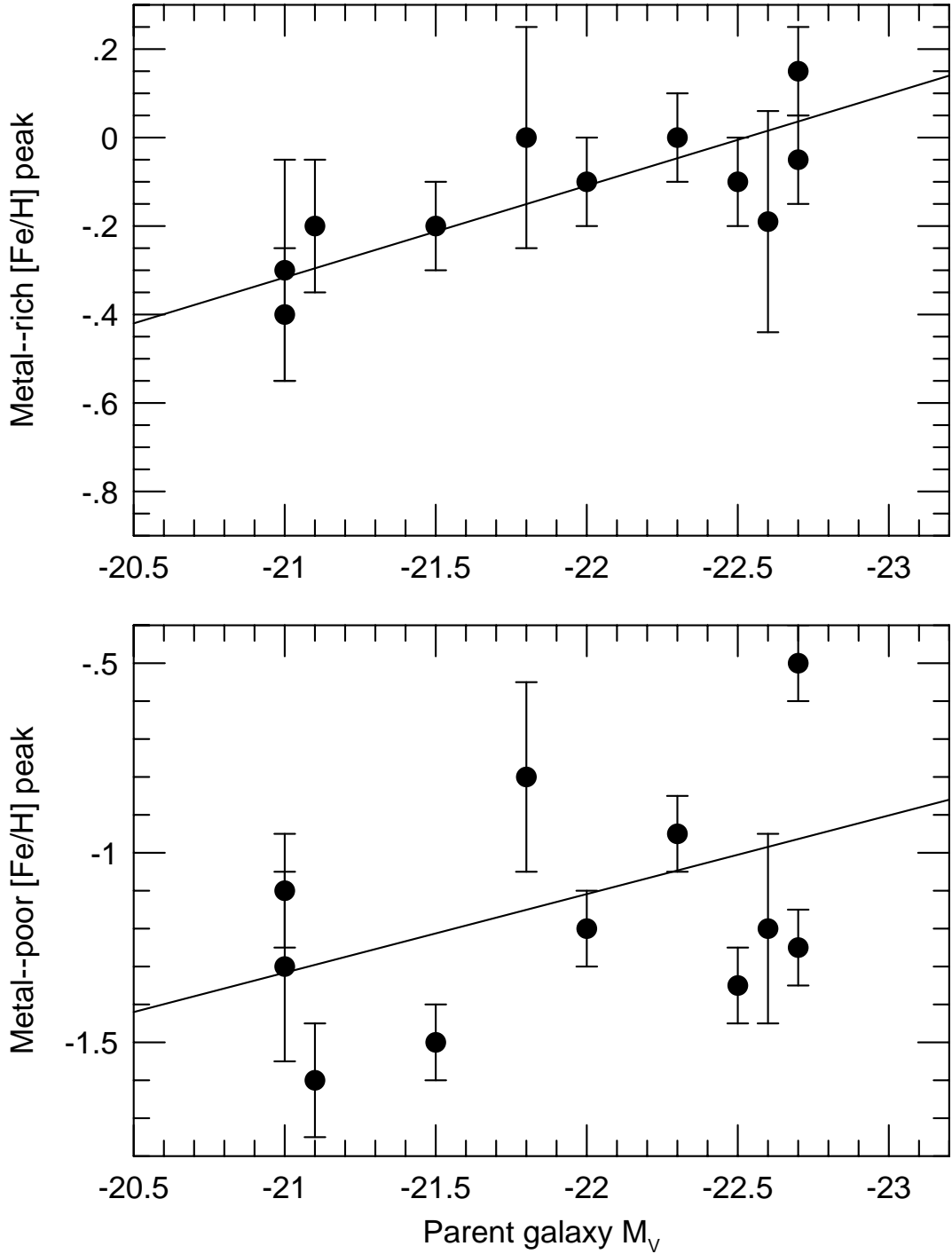}}
\caption{\label{fig1}
Globular cluster mean metallicity -- galaxy luminosity relation. The upper
panel shows the mean metallicity of the metal--rich population from
Table 1. 
The solid line is a weighted fit to the
data. A good correlation between the metallicity of the metal--rich
globular clusters and the parent galaxy luminosity is present. 
The lower panel shows the mean metallicity of the metal--poor 
population. The solid line is the metal--rich relation offset lower by
1 dex in [Fe/H] (i.e the average offset between the metal--rich and
metal--poor populations). There is little or no correlation between
the metallicity of the metal--poor 
globular clusters and the parent galaxy luminosity. 
}
\end{figure*}

\begin{figure*}[p]
\centerline{\psfig{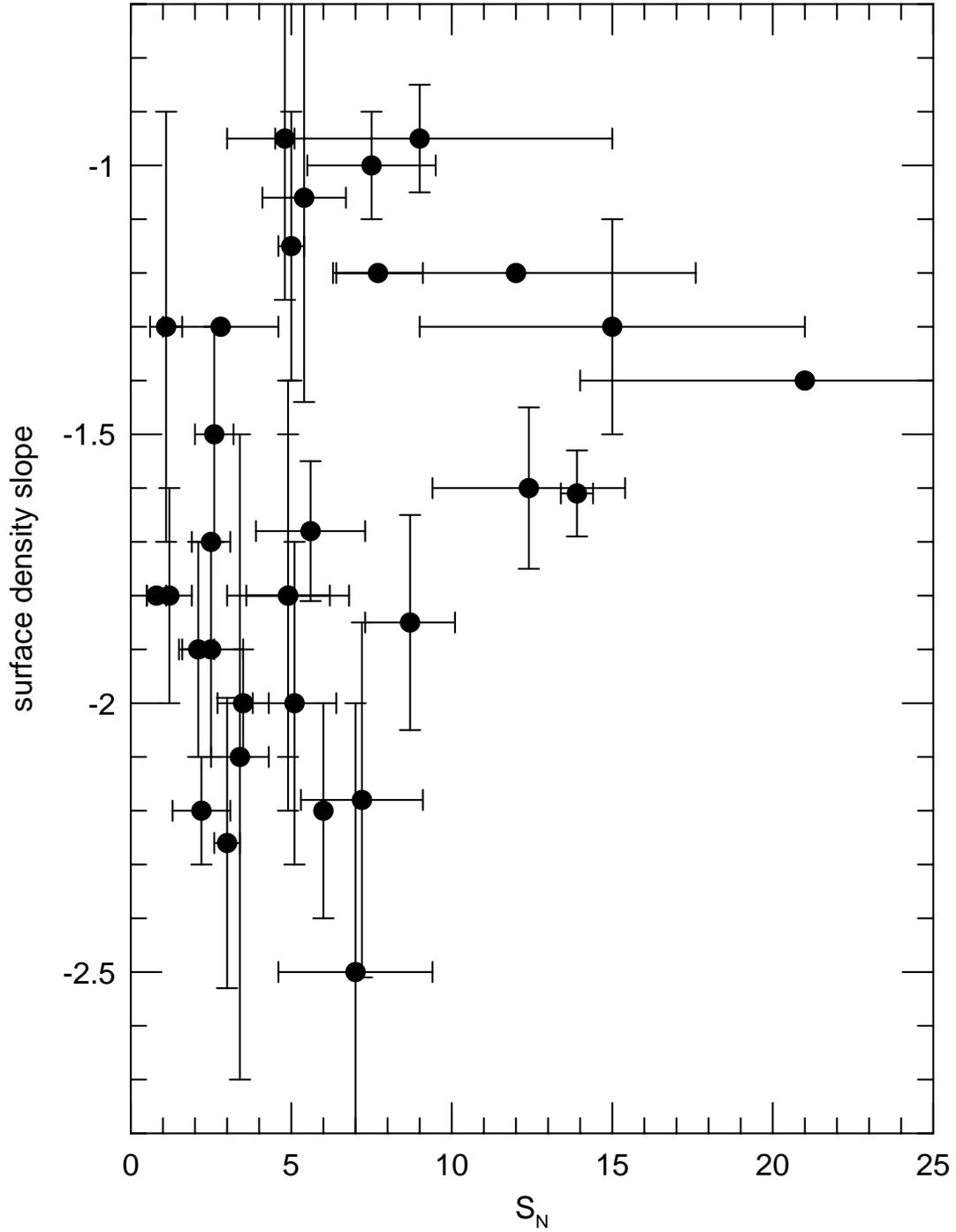}}
\caption{\label{fig2}
Globular cluster surface density slope versus specific frequency for
elliptical galaxies. The power--law slope of the globular cluster
surface density profile and the specific frequency (S$_N$) are taken from the
list of Kissler--Patig (1996). For low S$_N$ galaxies there is a wide
range of density slopes, but high S$_N$ galaxies only have flat,
extended globular cluster distributions. 
}
\end{figure*}

\begin{figure*}[p]
\centerline{\psfig{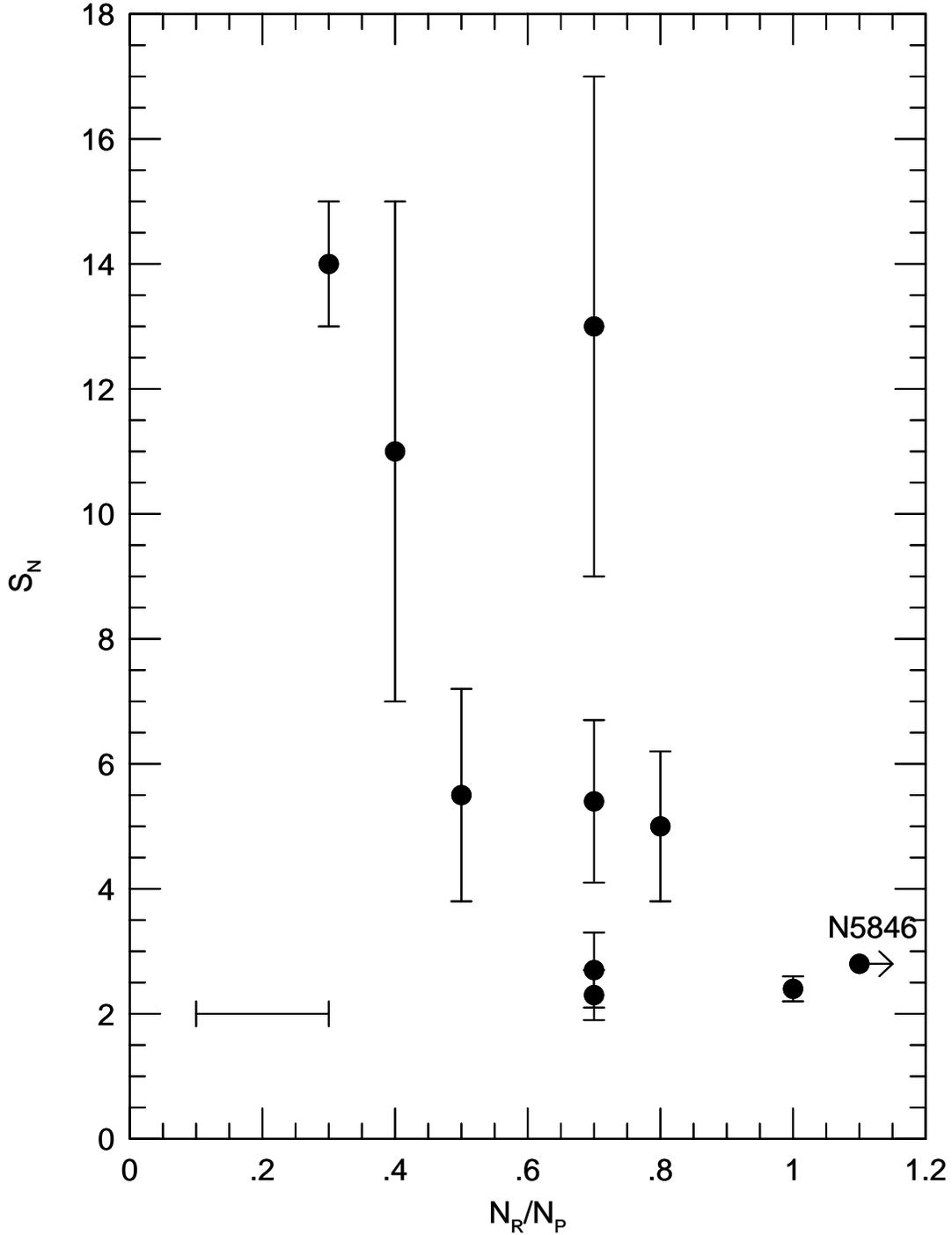}}
\caption{\label{fig3}
Specific frequency versus the ratio of metal--rich to metal--poor 
globular clusters. The data point for NGC 5846 (N$_R$/N$_P$ = 3.0,
S$_N$ = 2.8 $\pm$ 1.7) is not shown for display purposes. A typical
error bar for N$_R$/N$_P$ is shown in the lower left. 
The data show that as the ratio of metal--rich to metal--poor globular
clusters increases the specific frequency (S$_N$) value decreases. 
High S$_N$ galaxies 
have more metal--poor globular clusters than low S$_N$ ones. The
merger model of Ashman \& Zepf (1992) predicts an opposite trend
to that seen i.e., an increasing N$_R$/N$_P$ ratio with higher S$_N$
values. 
}
\end{figure*}

\begin{figure*}[p]
\centerline{\psfig{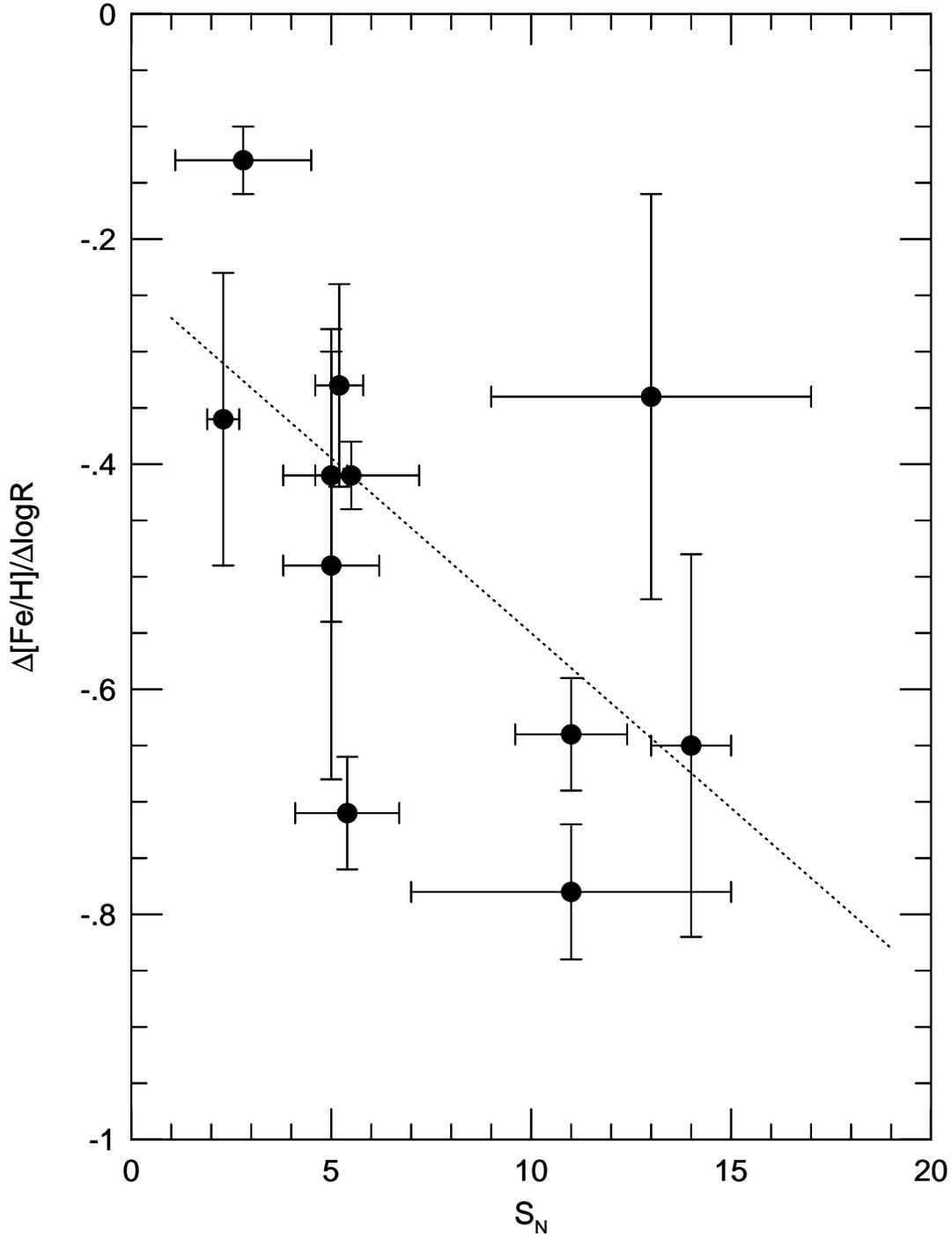}}
\caption{\label{fig4}
Globular cluster radial metallicity gradient versus specific
frequency. The data suggest that high S$_N$ galaxies (from Table 2) 
have steeper radial metallicity slopes than low S$_N$ ones. The dashed line
is the expected trend if an increase in S$_N$ corresponds to only larger
numbers of metal--poor globular clusters which are placed with the
same radial distribution as seen in NGC 4472 (see text for details). 
The zeropoint is taken to be the current values for NGC 4472
(i.e. S$_N$ = 5.5 and $\Delta$[Fe/H]/$\Delta$logR = --0.41). 
The merger model of Ashman \& Zepf (1992) predicts an opposite trend
to that seen i.e., shallower metallicity gradients in high S$_N$
galaxies.  
}
\end{figure*}

\begin{figure*}[p]
\centerline{\psfig{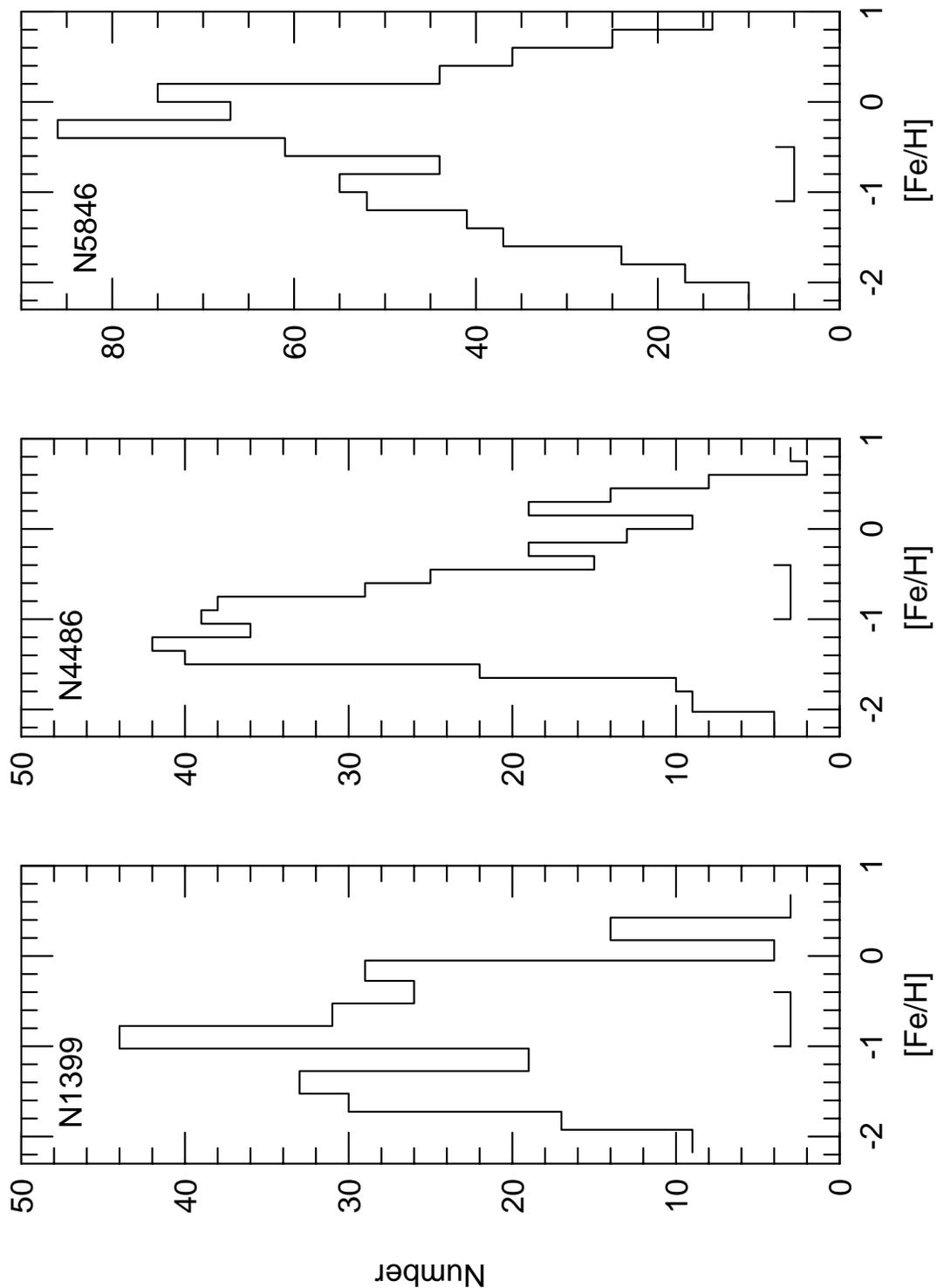}}
\caption{\label{fig5}
Globular cluster metallicity distribution in three central dominant
galaxies. The photometric data from Ostrov \etal (1993) for NGC 1399,
Giesler \etal (1996) for NGC 4486 and Forbes, Brodie \& Huchra 
(1996) for NGC 5846
have been converted into metallicity. The expected mean metallicity of
tidally stripped globular clusters from NGC 1404, NGC 4486B and NGC
5846A are marked below each distribution. In each case some of the
intermediate metallicity 
globular clusters may have been acquired by tidal stripping.  
}
\end{figure*}

\begin{figure*}[p]
\centerline{\psfig{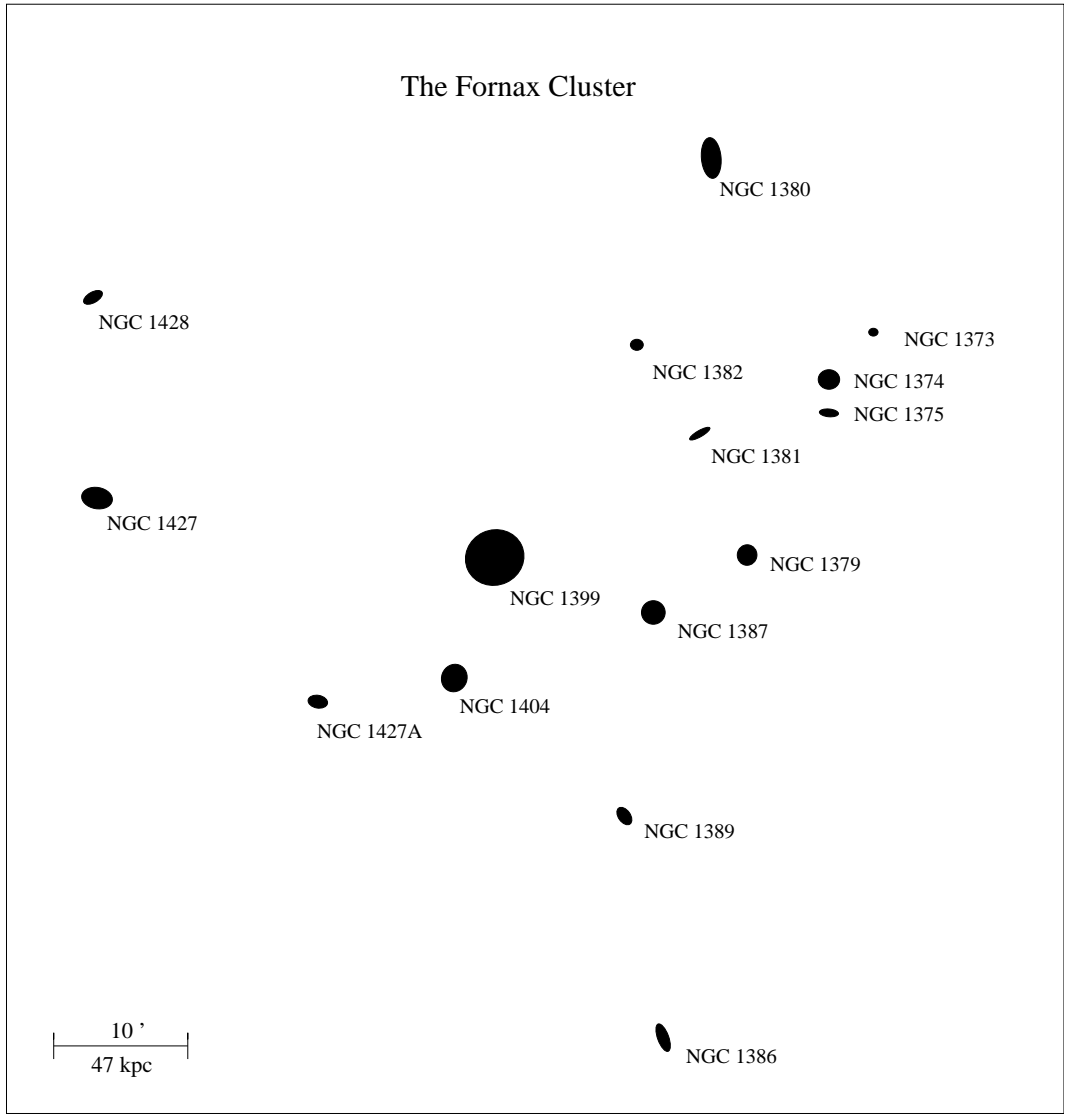}}
\caption{\label{fig6}
Schematic sketch of the galaxies in the central regions of the Fornax
cluster. The X--ray emission extends from the cD galaxy (NGC 1399) out
to at least 38$^{'}$. 
}
\end{figure*}

\begin{figure*}
\centerline{\psfig{figure=table1.epsi,width=450pt}}
%\caption{\label{}
%}
\end{figure*}

\begin{figure*}
\centerline{\psfig{figure=table2.epsi,width=450pt}}
%\caption{\label{}
%}
\end{figure*}

\begin{figure*}
\centerline{\psfig{figure=table3.epsi,width=450pt}}
%\caption{\label{}
%}
\end{figure*}

\begin{figure*}
\centerline{\psfig{figure=table4.epsi,width=450pt}}
%\caption{\label{}
%}
\end{figure*}

\end{document}